\documentclass{IEEEtran}
\usepackage{frank_defs}

\usepackage[normalem]{ulem}

\newcommand{\revised}[1]{%
\ifx\highlightrevisions\undefined{#1}%
\else\textcolor{red}{#1}%
\fi}

\newcommand{\revisedtwo}[1]{%
\ifx\highlightrevisionstwo\undefined{#1}%
\else\textcolor{red}{#1}%
\fi}

\newcommand{\revisedthree}[1]{%
\ifx\highlightrevisionsthree\undefined{#1}%
\else\textcolor{red}{#1}%
\fi}

\newcommand{\rnum}[1]{%
\ifx\showreviewercommentnum\undefined%
\else{[\bf{#1}] }%
\fi}

\newcommand{\del}[1]{%
\ifx\showerased\undefined%
\else{\sout{#1}}%
\fi}

\usepackage{amsthm,amsfonts,amssymb}
\usepackage{algpseudocode}
\usepackage[ruled,vlined]{algorithm2e}
\usepackage{array}
\usepackage{bbm}
\usepackage{bm}
\usepackage{float}
\usepackage{helvet}
\usepackage[colorlinks=true,linkcolor=blue,citecolor=magenta]{hyperref}
\usepackage{physics}
\usepackage{graphicx}
\usepackage{multirow}
\usepackage{mathtools}
\usepackage{url}
\usepackage{relsize}
\usepackage{tabularx}

\usepackage{newfloat}
\DeclareFloatingEnvironment[name={Supporting Figure}]{suppfigure}
 
\theoremstyle{definition}

\usepackage[url=false]{biblatex}

\begin{document} 

\begingroup
\let\clearpage\relax
\title{Accelerating Non-Cartesian MRI Reconstruction Convergence using k-space Preconditioning}
\author{Frank Ong, Martin Uecker, and Michael Lustig
\thanks{F. Ong (e-mail: franko@stanford.edu) is with the Department
of Electrical Engineering, Stanford University,
CA 94301, USA.}%
\thanks{M. Uecker (email: martin.uecker@med.uni-goettingen.de) is with the Department of Interventional and Diagnostic Radiology, University Medical Center Göttingen, 37075 Gottingen, Germany.}%
\thanks{M. Lustig (email: mlustig@eecs.berkeley.edu) is with the Department of Electrical Engineering and Computer Sciences, University of California, Berkeley, CA 94720, USA.}%
\thanks{This work was supported by NIH grant R01EB009690, Sloan Research Fellowship, Bakar Fellowship, and research support from GE Healthcare.}
}
\maketitle
\section{Abstract}
\label{sec:abstract}

We propose a k-space preconditioning formulation for accelerating the convergence of iterative Magnetic Resonance Imaging (MRI) reconstructions from non-uniformly sampled k-space data. Existing methods either use sampling density compensations which sacrifice reconstruction accuracy, or circulant preconditioners which increase per-iteration computation. Our approach overcomes both shortcomings. Concretely, we show that viewing the reconstruction problem in the dual formulation allows us to precondition in k-space using density-compensation-like operations. Using the primal-dual hybrid gradient method, the proposed preconditioning method does not have inner loops and are competitive in accelerating convergence compared to existing algorithms. We derive $\ell 2$-optimized preconditioners, and demonstrate through experiments that the proposed method converges in about ten iterations in practice.

\begin{IEEEkeywords}
MRI, Iterative Reconstruction, Non-Cartesian, Preconditioner, Density Compensation
\end{IEEEkeywords}

\section{Introduction}

Non-Cartesian trajectories provides many advantages over Cartesian sampling based on their unique properties. Spiral~\cite{ahn1986high, meyer1992fast} and cones trajectories~\cite{gurney2006design}, for example, can be designed to traverse k-space efficiently, which make them suitable for fast imaging applications, including coronary imaging~\cite{meyer1992fast}, and arterial spin labeled perfusion imaging~\cite{alsop2015recommended}. Many non-Cartesian trajectories, such as radial~\cite{lauterbur1973image} and projection reconstruction~\cite{glover1992projection} naturally sample low-frequency regions densely, which can provide auto-calibration regions for parallel imaging (PI), and robustness to motion for dynamic applications. Such variable density sampling~\cite{tsai2000} property is also more adapted to signal energy than uniform sampling, which results in less coherent undersampling artifacts in the wavelet transform domain. Hence, variable density non-Cartesian trajectories are often used with compressed sensing (CS)~\cite{Lustig:2007cu}.

On the other hand, reconstructions from non-Cartesian trajectories, especially with PI, are more complex and time-consuming than from Cartesian trajectories. The long reconstruction time is one reason that has limited the clinical adoption of non-Cartesian trajectories. Iterative reconstructions, such as CG-SENSE~\cite{pruessmann2001}, have to be used for PI, which can often take many iterations to converge. In comparison, the Cartesian SENSE method~\cite{pruessmann1999sense} has an analytic solution that can be efficiently solved in a single step.

\begin{figure}[!ht]
\begin{center}
  \includegraphics[width=\linewidth]{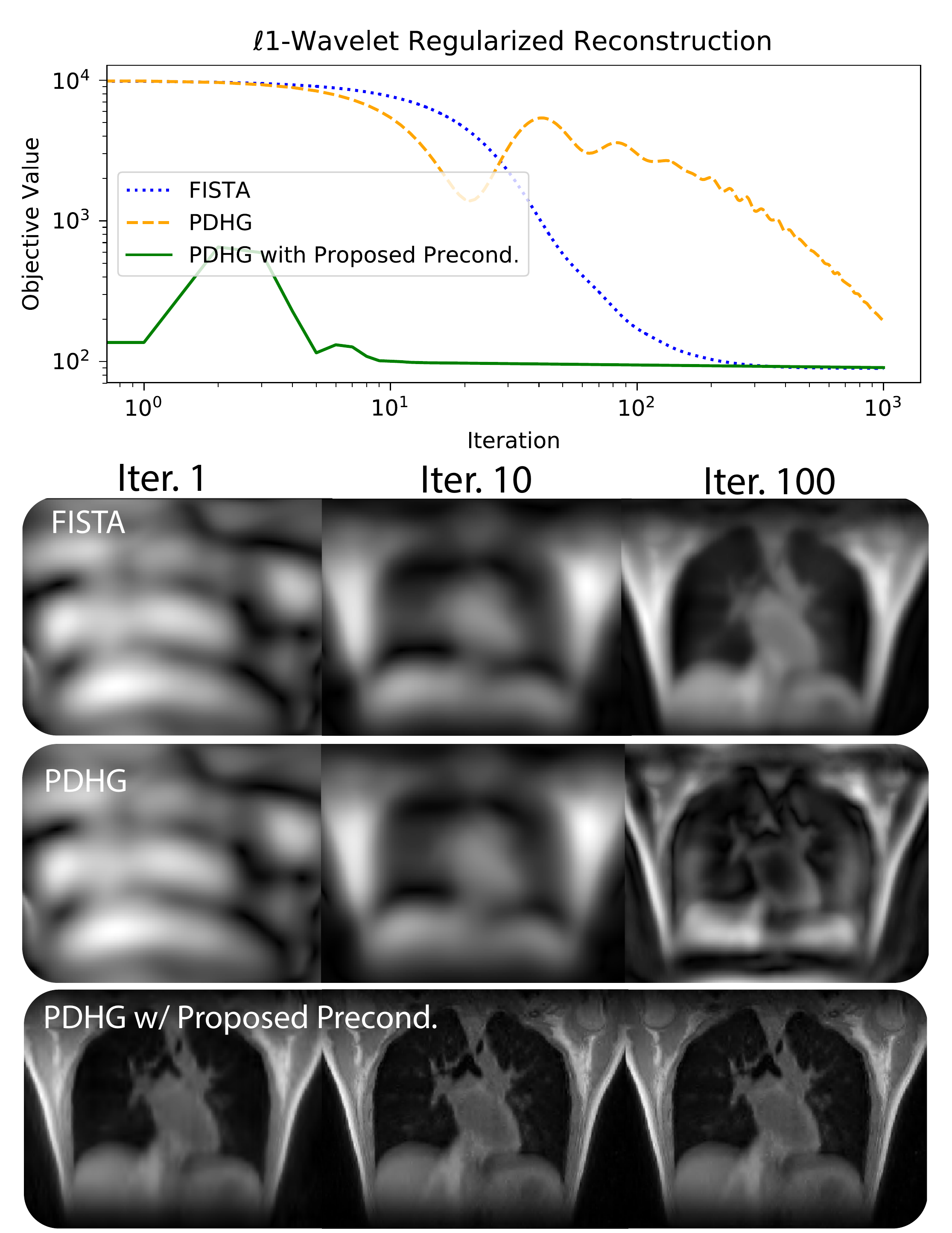}
\caption{Iteration progression for $\ell 1$ wavelet regularized reconstruction of a 3D UTE lung dataset. Both FISTA and PDHG exhibit extreme blurring even after 100 iterations. In contrast, PDHG with the proposed preconditioner (multi-channel k-space preconditioner) converges in about ten iterations, both visually and quantitatively in terms of minimizing the objective value.}
\label{fig:l1_wavelet_uwute}
\end{center}
\end{figure}

One way to make non-Cartesian PI/CS reconstructions more efficient is reducing the number of iterations. In general, the slow convergence of iterative methods is due to the ill-conditioning of the reconstruction problem. For non-Cartesian imaging, such ill-conditioning comes from the variable density sampling distribution in k-space. This often shows up in images as blurring artifacts when the reconstruction has not yet converged. Slow convergence is even more significant for 3D acquisitions and CS reconstructions. For instance, Figure~\ref{fig:l1_wavelet_uwute} shows the iteration progression for a $\ell 1$-wavelet regularized reconstruction of a 3D ultra-short echo-time (UTE) radial acquisition using the Fast Iterative Soft-Thresholding Algorithm (FISTA)~\cite{beck2009}, and primal dual hybrid gradient method (PDHG)~\cite{chambolle2011} (also known as the Chambolle-Pock method). Even after 100 iterations, the reconstructed image still displays significant blurring due to slow convergence.

Density compensation~\cite{jackson1991, meyer1992fast, hoge1997, pipe1999} is often used as a heuristic to compensate for slow convergence in non-Cartesian iterative reconstruction. It was originally developed for gridding reconstruction, and was mostly designed for Nyquist-sampled trajectories. The use of density compensation in iterative PI reconstruction was first introduced by Pruessmann et al.~\cite{pruessmann2001}. While their work showed that in practice density compensation can speed up convergence, reconstruction error was also increased. This is because the data consistency for densely sampled regions is weighted down in the objective function (more detail in Section~\ref{sec:dcf}).

An alternative to density compensation is preconditioning. Preconditioning has the advantage of preserving the original objective function and hence does not affect the reconstruction accuracy. Many techniques~\cite{sutton2003, ramani2011, weller2014, muckley2016, koolstra2018, trzasko2014} have been proposed for MRI iterative reconstruction as described in detail in Section~\ref{sec:prior}. However, a drawback of existing methods is that they increase the per-iteration computation. In particular, most existing preconditioners have circulant structures, and require at least two additional FFTs per iteration. Moreover, all prior methods require inner loops in their algorithms for non-Cartesian reconstructions, \revised{\rnum{R1.1}which result in more parameters to tune and can often incur additional computational overhead from initializing inner loop variables.}

In this article, we present a framework for speeding up convergence that combines the computational efficiency of density compensation, and the objective preserving property of preconditioning. Similar to the work of Trzasko et al.~\cite{trzasko2014}, we consider using efficient operations in k-space for preconditioning. Our contribution is to recognize that a diagonal preconditioner can be applied in k-space more generally by viewing the objective function in the dual formulation. Using PDHG~\cite{chambolle2011}, the resulting method with preconditioning does not have inner loops, so it has a similar computational complexity as the vanilla proximal gradient method. Moreover, instead of using off-the-shelf density compensation factors, we derive $\ell 2$-optimized diagonal preconditioners for the MRI forward model. We demonstrate through experiments that the proposed diagonal preconditioner speeds up iterative reconstruction for non-Cartesian imaging, with $\ell 2$, $\ell 1$-wavelet, and total variation regularizations.

\section{Problem Setup}
\label{sec:problem}

Throughout this article, we consider the following discrete multi-channel MRI forward model, in which we are given an $N$-size image  $\mathbf{x} \in \mathbb{C}^N$, $C$-channel sensitivity maps $\{\mathbf{s}_c \in \mathbb{C}^{N}\}_{c = 1}^C$,  white Gaussian noise vectors $\{\mathbf{w}_c \in \mathbb{C}^{M}\}_{c = 1}^C$, and k-space measurements $\{\mathbf{y}_c \in \mathbb{C}^{M}\}_{c = 1}^C$ with sampling points  $\{ f_i \}_{i = 1}^M$ such that
\begin{align}
  \label{eq:model_i}
  \mathbf{y}_{c}[i] = \frac{1}{\sqrt{N}} \sum_{n = 0}^{N - 1} \mathbf{s}_{c}[n] \mathbf{x}[n] e^{-\imath 2 \pi f_i n/N} + \mathbf{w}_{c}[i]
\end{align}
for $i \in \{1, \ldots, M\}$, and $c \in \{1, \ldots, C\}$. For notation simplicity, we focus on one-dimensional signals. The above model can be succinctly represented as a linear model:
\begin{align}
  \label{eq:model}
  \mathbf{y} = \mathbf{A} \mathbf{x} + \mathbf{w}
\end{align}
where $\mathbf{y} \in \mathbb{C}^{MC}$ and $\mathbf{w} \in \mathbb{C}^{MC}$ are stacked versions of $\{\mathbf{y}_c \in \mathbb{C}^{M}\}_{c = 1}^C$ and $\{\mathbf{w}_c \in \mathbb{C}^{M}\}_{c = 1}^C$.

Given the acquired k-space measurements $\mathbf{y}$, we consider the following regularized least squares problem to reconstruct the image:
\begin{align}
  \label{eq:obj}
  \min_{\mathbf{x}} \frac{1}{2} \| \mathbf{A} \mathbf{x} - \mathbf{y} \|_2^2 + g(\mathbf{x})
\end{align}
where $g(\mathbf{x})$ is the regularization function. \revised{\rnum{R1.3}Specifically, we will consider three regularization functions as concrete examples: $\ell 2$-norm $(\lambda \| \mathbf{x} \|_2^2 / 2)$, $\ell 1$-wavelet $(\lambda \| \mathbf{W} \mathbf{x} \|_1)$, where $\mathbf{W} \in \mathbb{C}^{N \times N}$ is a unitary wavelet transform operator, and total variation $(\lambda \| \mathbf{G} \mathbf{x} \|_1)$, where $\mathbf{G} \in \mathbb{C}^{N \times N}$ is a first-order finite difference operator with periodic boundary extension.}

Since the image size $N$ is on the order of tens of thousands or more, the above reconstruction problem is in practice only solved approximately using first-order gradient methods. In the following section, we will focus on the proximal gradient method as an example to illustrate the advantages and disadvantages of using density compensation and preconditioners to accelerate convergence.

The proximal gradient method when applied to objective function~\eqref{eq:obj} gives the following update for the $k$th iteration:
\begin{align}
  \label{eq:prox_grad}
  \mathbf{x}^{(k + 1)} = \prox_{\alpha g}(\mathbf{x}^{(k)} - \alpha \mathbf{A}^H (\mathbf{A} \mathbf{x}^{(k)} - \mathbf{y}))
\end{align}
where $\prox_{\alpha g}(\mathbf{z}) = \text{argmin}_\mathbf{x} \| \mathbf{x} - \mathbf{z} \|_2^2 / (2 \alpha) + g(\mathbf{x})$.

The convergence rate depends only on $\mathbf{A}^H \mathbf{A}$. More concretely, when $\mathbf{A}$ is not singular, then the step-size $\alpha$ can be chosen so that the convergence rate is inversely proportional to the condition number of $\mathbf{A}^H \mathbf{A}$. When $\mathbf{A}$ is singular, then the step-size can be chosen so that the convergence rate is inversely proportional to the maximum eigenvalue of $\mathbf{A}^H \mathbf{A}$. For variable density sampling, the condition number or the maximum eigenvalue of $\mathbf{A}^H \mathbf{A}$ is much higher than for uniform density sampling for a given undersampling factor and hence results in slow convergence.

\section{\revised{\rnum{R1.2}Prior Arts on Density Compensation and Image Domain Preconditioning}}
\label{sec:prior}

\subsection{Density Compensation}
\label{sec:dcf}

One effective heuristic to accelerate convergence for non-Cartesian imaging is incorporating density compensation factors during iterations. Given a diagonal matrix $\mathbf{D} \in \mathbb{C}^{MC \times MC}$ with density compensation factor as diagonals, the heuristic modifies the proximal gradient method as follows:
\begin{align}
  \label{eq:dcf_prox_grad}
  \mathbf{x}^{(k + 1)} = \prox_{\alpha g}(\mathbf{x}^{(k)} - \alpha \mathbf{A}^H \mathbf{D} (\mathbf{A} \mathbf{x}^{(k)} - \mathbf{y}))
\end{align}

The use of density compensation in iterative PI reconstruction was first introduced by Pruessmann et al.~\cite{pruessmann2001}, and was shown to speed up convergence in practice. Computationally, incorporating density compensation in each iteration costs an additional $O(M C)$ multiplications, adding very little overhead to the overall iteration. However, the main drawback is that such k-space weighting is known to increase reconstruction errors, as implicitly it is solving for a weighted objective function:
\begin{align}
  \label{eq:dcf_obj}
  \min_{\mathbf{x}} \frac{1}{2} \| \mathbf{D}^{1/2} (\mathbf{A} \mathbf{x} - \mathbf{y}) \|_2^2 + g(\mathbf{x})
\end{align}
Note that data consistency is weighed down in densely sampled regions. Measurements are essentially thrown away for convergence, resulting in increased reconstruction error, and noise coloring.

\subsection{Image-domain Preconditioning}
\label{sec:image_domain_preconditioning}

An alternative is to use preconditioning, which only affects the convergence, but not the objective function. Since the objective function is not changed, there is no error penalty for using preconditioners. 

In particular, given a preconditioner $\mathbf{P} \in \mathbb{C}^{N \times N}$, the preconditioned proximal gradient method applies:
\begin{align}
  \label{eq:precond_prox_grad}
  \mathbf{x}^{(k + 1)} = \prox_{\alpha g, \mathbf{P}}(\mathbf{x}^{(k)} - \alpha \mathbf{P} \mathbf{A}^H (\mathbf{A} \mathbf{x}^{(k)} - \mathbf{y}))
\end{align}
The preconditioner $\mathbf{P}$ should be designed to approximate the (pseudo) inverse of $\mathbf{A}^H \mathbf{A}$ such that the condition number or maximum eigenvalue of $\mathbf{P} \mathbf{A}^H \mathbf{A}$ is much lower than that of $\mathbf{A}^H \mathbf{A}$. 

Many works have proposed the use of preconditioning to accelerate MRI reconstructions. In particular, it was first described by Sutton et al.~\cite{sutton2003} for single-channel non-Cartesian imaging in the presence of field inhomogeneities. It was further explored by Ramani et al.~\cite{ramani2011} for PI-CS reconstructions. Their method leveraged a circulant preconditioner developed by Yagle~\cite{yagle2002} for Toeplitz systems. Weller et al.~\cite{weller2014} considered the non-Cartesian $\ell 1$-SPIRiT~\cite{lustig2010} method and used an $\ell 2$-optimal circulant preconditioner developed by Chan~\cite{chan1988}. Muckley et al.~\cite{muckley2016} considered FISTA~\cite{beck2009} and designed a circulant preconditioner that majorizes the sensing matrix motivated by the convergence criterion. Finally, Koolstra et al.~\cite{koolstra2018} considered the split-Bregman method for Cartesian PI-CS reconstructions and presented a circulant preconditioner that incorporates multi-channel sensitivity maps in the construction of their proposed preconditioner.

A main drawback of the above mentioned preconditioning methods is that they all increase per-iteration computational complexity. This is because to compensate ill conditioning from variable density in k-space, existing preconditioners have to use circulant operators, which cost two FFTs per iteration. That is, existing preconditioners are of the form,
\begin{align}
    \mathbf{P} = \mathbf{F} \text{diag}(\mathbf{p}) \mathbf{F}^H
\end{align}
where $\mathbf{p} \in \mathbb{C}^N$ is a Fourier weighting vector, and $\mathbf{F} \in \mathbb{C}^{N \times N}$ is the unitary discrete Fourier transform operator.

A more subtle issue is that all of them require inner loops in their algorithms when incorporating CS with non-Cartesian MRI. In particular, the proximal operator has to be modified to incorporate the preconditioner, which requires inner iterations to solve even when the proximal operator is simple:
\begin{align}
  \prox_{\alpha g, \mathbf{P}}(\mathbf{z}) = \underset{\mathbf{x}}{\text{argmin}} \frac{1}{2 \alpha} \| \mathbf{P}^{-1/2} \mathbf{x} - \mathbf{z} \|_2^2 + g(\mathbf{x}).
\end{align}

In summary, although existing preconditioners have shown that they can accelerated convergence, their shortcoming lies in the per-iteration increase in complexity.

\section{Diagonal k-space Preconditioning}
\label{sec:fourier_precond}

Ideally, we would like to develop a preconditioning method that can achieve the computational efficiency of density compensation without changing the objective function. 

Recently, Trzasko et al.~\cite{trzasko2014} showed that through an algebraic manipulation, a diagonal preconditioner can be applied in k-space for the least squares sub-problem of ADMM~\cite{Boyd:2011bw}. This enables a different mechanism for preconditioning. In particular, they show that it is possible to use efficient operations in k-space for preconditioning. However, their formulation still required inner loops to solve for the sub-problem.

Here we show that k-space preconditioning without inner loops is achievable by looking at the convex dual problem. Since the reconstruction problem~\eqref{eq:obj} is unconstrained, it must satisfy strong duality. Its corresponding dual problem (see the Supporting Materials for a derivation using the Lagrangian) is given by:
\begin{align*}
  \max_\mathbf{u} -\left(\frac{1}{2}\| \mathbf{u} \|_2^2 - \Re \langle \mathbf{u}, \mathbf{y} \rangle + g^*(-\mathbf{A}^H \mathbf{u}) \right)
\end{align*}
where $\mathbf{u} \in \mathbb{C}^m$ is the dual variable and $g^*$ denotes the convex conjugate function of $g$. 

Our key observation is that because the dual variable resides in k-space, it is now possible to perform preconditioning in k-space on the dual problem.

In general, one would require primal-dual methods to solve for the primal and dual problems at the same time. In this article, we opt for the PDHG~\cite{chambolle2011} method for an algorithm without inner loops. Other primal-dual reconstruction methods, such as those described in the work of Komodakis et al.~\cite{komodakis2015}, can also be used.

Before describing the algorithm, we note that the $\ell 2$-regularized reconstruction is a special case that can efficiently recover the primal variable from the dual problem. While this property is not used in our experiments, we show that this is connected to Trzasko et al.~\cite{trzasko2014} in the Supporting Materials.

\subsection{Preconditioned PDHG for simple regularizers}
\label{sec:l1}
  
\revised{Let us first consider the case when the regularizer $g(\mathbf{x})$ is simple, i.e. its proximal operator is easy to compute. An example is the $\ell 1$ wavelet regularization function $g(\mathbf{x}) = \lambda \| \mathbf{W} \mathbf{x} \|_1$, where $\mathbf{W}$ is the wavelet transform operator.} 

Following~\cite{chambolle2011} and~\cite{pock2011}, for each iteration $k$, the diagonal preconditioned version of PDHG for simple proximal operators with a diagonal preconditioner $\mathbf{P} \in \mathbb{C}^{MC \times MC}$ is given by, 
\begin{align*}
  \mathbf{u}^{(k + 1)} &= (\mathbf{I} + \sigma^{(k)} \mathbf{P})^{-1} (\mathbf{u}^{(k)} + \sigma^{(k)} \mathbf{P} (\mathbf{A} \bar{\mathbf{x}}^{(k)} - \mathbf{y}))\\
  \mathbf{x}^{(k + 1)} &= \prox_{\tau^{(k)} g} (\mathbf{x}^{(k)} - \tau^{(k)} \mathbf{A}^H \mathbf{u}^{(k + 1)})\\
  \bar{\mathbf{x}}^{(k + 1)} &= \mathbf{x}^{(k + 1)} + \theta^{(k)} (\mathbf{x}^{(k + 1)} - \mathbf{x}^{(k)}),
\end{align*}
where $\bar{\mathbf{x}}^k$ and $\theta^k$  are the extrapolated primal variable and extrapolation parameter to provide acceleration, and $\tau^k$ and $\sigma^k$ are the primal and dual step-size respectively such that
\begin{align*}
  \sigma^{(k)} \tau^{(k)} \lambda_{\text{max}}( \mathbf{P} \mathbf{A} \mathbf{A}^H ) < 1,
\end{align*}
where $\lambda_{\text{max}}$ denotes the maximum eigenvalue. 

\revised{\rnum{R1.6, R3.2}One drawback of primal-dual methods is that there are many ways of choosing the step-sizes, which can affect the convergence rate. In this article, we adopt a simple scheme, which reflects that in practice it is impossible to fine-tune parameters to each dataset. We fix $\sigma^{(0)} = 1$ and choose $\tau^{(0)} = 1 / \lambda_{\text{max}}( \mathbf{P} \mathbf{A} \mathbf{A}^H )$. Since the problem is dual strongly convex~\cite{chambolle2011} (the function involving the dual variable $\mathbf{u}$ is the $\ell 2$ norm), acceleration can be obtained by choosing the subsequent step-sizes appropriately as:}
\begin{align*}
  \theta^{(k)} &= 1 / (1 + 2 \sigma^{(k)} \min_i \mathbf{p}_i) \\
  \sigma^{(k + 1)} &= \theta^{(k)} \sigma^{(k)} \\
  \tau^{(k + 1)} &= \tau^{(k)} / \theta^{(k)},
\end{align*}
which can be derived from Theorem 5.1 in \cite{chambolle_stochastic_2018} with $\mu_i = \min_j \mathbf{p}_j / \mathbf{p}_i$ and $\tilde \sigma_{(0)} = \min_j \mathbf{p}_j$.

\subsection{Preconditioned PDHG for simple regularizers composed with linear operators}
\label{sec:tv}

Next, we consider regularization functions which consist of a simple function $h$ composed with a linear operator $\mathbf{G}$:
\begin{align*}
    g(\mathbf{x}) = h(\mathbf{G} \mathbf{x}).
\end{align*}
One example is the total variation regularization $g(\mathbf{x}) = \lambda \| \mathbf{G} \mathbf{x} \|_1$ with $\mathbf{G}$ being the finite-difference operator.

In this case, PDHG can be modified to perform:
\begin{align*}
  \mathbf{u}^{(k + 1)} &= (\mathbf{I} + \sigma^{(k)} \mathbf{P})^{-1} (\mathbf{u}^{(k)} + \sigma^{(k)} \mathbf{P} (\mathbf{A} \bar{\mathbf{x}}^{(k)} - \mathbf{y}))\\
  \mathbf{v}^{(k + 1)} &= \prox_{\tau^{(k)} h} (\mathbf{v}^k + \sigma^{(k)} \mathbf{G} \bar{\mathbf{x}}^{(k)})\\
  \mathbf{x}^{(k + 1)} &= \mathbf{x}^{(k)} - \tau^{(k)} (\mathbf{A}^H \mathbf{u}^{(k + 1)} + \mathbf{G}^H \mathbf{v}^{(k + 1)})\\
  \bar{\mathbf{x}}^{k + 1} &= \mathbf{x}^{(k + 1)} + \theta^{(k)} (\mathbf{x}^{(k + 1)} - \mathbf{x}^{(k)}),
\end{align*}
where $\tau^{(k)}$ and $\sigma^{(k)}$ are the primal and dual step-sizes respectively such that
\begin{align*}
  \sigma^{(k)} \tau^{(k)} (\lambda_{\text{max}}( \mathbf{P} \mathbf{A} \mathbf{A}^H) + \lambda_{\text{max}}(\mathbf{G} \mathbf{G}^H )) < 1.
\end{align*}

\revised{Note that unlike the previous case, the problem is no longer dual strongly convex (the function involving the dual variable $\mathbf{v}$ is $h$, which in general is not strongly convex), so the same acceleration scheme cannot be used. We choose $\theta^{(k)} = 1$, $\sigma^{(k)} = 1$ and $\tau^{(k)} = 1 / (\lambda_{\text{max}}( \mathbf{P} \mathbf{A} \mathbf{A}^H) + \lambda_{\text{max}}(\mathbf{G} \mathbf{G}^H))$ for all $k$.}

\section{$\ell 2$ optimized diagonal k-space preconditioners}
\label{sec:precond}

Now that we know how to precondition in k-space, it becomes clear from the dual problem that the preconditioner should be designed to precondition the matrix $\mathbf{A} \mathbf{A}^H$. In this article, we consider diagonal preconditioners to approximate the inverse of the normal operator $\mathbf{A} \mathbf{A}^H$ in the least squares sense. The diagonal structure is desired because we want to apply the preconditioner efficiently in k-space, similarly to density compensation. The least squares design, on the other hand, is used here so that we can efficiently compute the preconditioner.

Concretely, we consider a diagonal preconditioner $\mathbf{p} \in \mathbb{C}^{MC}$ such that,
\begin{align*}
  \mathbf{p} = \underset{\mathbf{p}}{\text{argmin}} \left\|  \text{diag}(\mathbf{p}) \mathbf{A} \mathbf{A}^H - \mathbf{I} \right\|_F^2.
\end{align*}

Let $\mathbf{a}_i \in \mathbb{C}^N$ denote the $i$th row vector of $\mathbf{A}$. As shown in Appendix~\ref{sec:appendix_precond}, the general expression for the inverse of the diagonal preconditioner is given by:
\begin{align*}
  {\mathbf{p}[i]}^{-1} = \frac{ \sum_{j=1}^M |\mathbf{a}_i^H \mathbf{a}_j|^2 } { \|\mathbf{a}_i\|_2^2 }.
\end{align*}

To further look into the preconditioner, we first consider the single-channel case. In this case, $\mathbf{a}_{i}[n] = e^{-\imath 2 \pi f_i n / N} / \sqrt{N}$, and $\| \mathbf{a}_i \|_2^2 = 1$. Then the diagonal preconditioner at k-space position $i$, is given by:
\begin{equation}
\begin{aligned}
 {\mathbf{p}[i]}^{-1}  &=  \frac{1}{N}  \sum_{j = 1}^M  \left| \sum_{n = 0}^{N - 1}  e^{-\imath 2 \pi (f_i - f_j) n/N} \right|^2 \\
&=  \frac{1}{N} \sum_{j = 1}^M \left| \frac{\sin(\pi (f_i - f_j))}{\sin(\pi (f_i-f_j) / N)} \right|^2
\end{aligned}
\label{eq:sc_precond}
\end{equation}

For Cartesian trajectories, the frequency spacing $f_i - f_j$ are all integers, and hence $\mathbf{p}[i] = 1$ for all $i$, which matches our expectation that the condition number for single channel Cartesian imaging systems cannot be improved. For non-Cartesian trajectories, the diagonal preconditioner can be interpreted as calculating density from the sinc squared kernel $| \sin(\pi f) / (N \sin(\pi f / N)) |^2$.

Moving on to multi-channel, for k-space position $i$ and coil $c$, the row vector is given by $\mathbf{a}_{ci}[n] = \mathbf{s}_{c}[n] e^{-\imath 2 \pi f_i n / N} / \sqrt{N}$. Hence, we obtain,
\begin{equation}
  {\mathbf{p}_c[i]}^{-1} =  \frac{1}{\| \mathbf{s}_c \|_2^2 N} \sum_{j = 1}^M \sum_{d = 1}^C
                  \left| \sum_{n = 0}^{N - 1}  \mathbf{s}_{c}[n] \mathbf{s}^*_{d}[n] e^{-\imath 2 \pi (f_i - f_j) n/N} \right|^2
\label{eq:mc_precond}
\end{equation}

Here we pause to note that the multi-channel preconditioner design is different from density compensation calculations in that we incorporate coil sensitivity maps. One downside is that the multi-channel preconditioner has to be calculated whenever the coil sensitivity maps and the sampling trajectory change. For many clinical applications, the coil sensitivity maps are computed from a pre-scan or estimated from the first scan and used multiple times for a sequence of scans. In this case, the overhead of computing the preconditioner becomes negligible. \revised{For applications in which this overhead matters, the single-channel preconditioner~\eqref{eq:sc_precond} can be used as an approximation. As shown in our experiments in Section~\ref{sec:results}, we found that the single-channel preconditioner is competitive in accelerating convergence.}

Since the multi-channel preconditioner has to be computed whenever the coil sensitivity maps change, its computation time should not be impractically long. A direct summation implementation takes $O(M^2 N C^2)$ computation. In the following, we show that using Fourier transform properties, we can reduce the computational complexity to $O(C^2 N \log N + C M)$, which makes it comparable to common calibration methods, such as ESPIRiT~\cite{uecker2014}. Figure~\ref{fig:diagram} provides a high-level diagram of the overall process.

\subsection{Efficient computation of the multi-channel preconditioner}

First, we note that we can express the squared terms for the multi-channel preconditioner with cross-correlations, which can be computed in $O(C^2 N \log N)$ using FFTs. Let us define,
\begin{align*}
  \mathbf{r}_{cd}[k] &= \sum_{m - n = k} (\mathbf{s}_{c}[n] \mathbf{s}^*_{d}[n])^* (\mathbf{s}_{c}[m] \mathbf{s}^*_{d}[m])
\end{align*}
Then
\begin{align*}
  \left| \sum_{n = 0}^{N - 1} \mathbf{s}_{c}[n] \mathbf{s}^*_{d}[n] e^{-\imath 2 \pi (f_i - f_j) n/N} \right|^2 \\= \sum_{k = - N + 1}^{N - 1} \mathbf{r}_{cd}[k] e^{-\imath 2 \pi (f_i - f_j) k /N}
\end{align*}

Next, we note that the preconditioner can be expressed in terms of convolution with the point spread function, which can be approximated using NUFFT with $O(N \log N + M)$ computational complexity. Let us define
\begin{align*}
  \mathbf{h}[k] &= \frac{1}{\sqrt{N}} \sum_{j = 1}^M e^{\imath 2 \pi f_j k/N},
  \end{align*}
 then,
\begin{align*}
  {\mathbf{p}_c[i]}^{-1} &= \frac{1}{\| \mathbf{s}_c \|_2^2 \sqrt{N}} \sum_{k = - N + 1}^{N - 1}  \sum_{d=1}^C \mathbf{r}_{cd}[k] \mathbf{h}[k] e^{-\imath 2 \pi f_i k/N}.
\end{align*}
The final step involves $C$ NUFFTs on the point-wise multiplication of $\mathbf{r}$, and $\mathbf{h}$. Hence we obtain the overall computational complexity to be $O(C^2 N \log N + C M)$.

\begin{figure}[H]
\begin{center}
  \includegraphics[width=\linewidth]{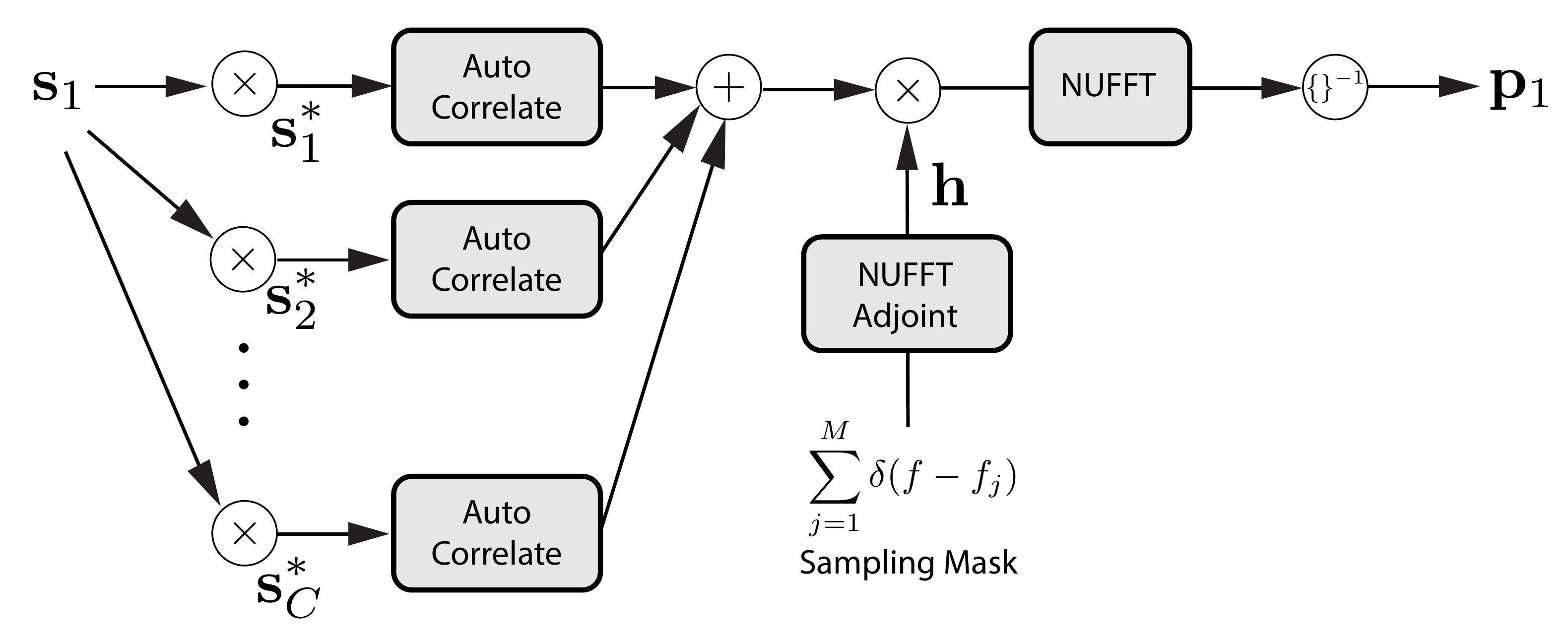}
\caption{Diagram of computing the multi-channel diagonal k-space preconditioner for the first channel. The same process can be used for the single-channel k-space preconditioner by setting $C=1$ and $\mathbf{s}_1$ as an all-one image.}
\label{fig:diagram}
\end{center}
\end{figure}

\section{Experiments}
\label{sec:experiments}

In the spirit of reproducible research, we provide a software package in Python to reproduce the results described in this article. The software package can be downloaded from:
\begin{center}
  \url{https://github.com/mikgroup/kspace_precond.git}
\end{center}

For each regularization, we evaluated the proposed method on three 2D non-Cartesian datasets: 1) a liver dataset acquired with a stack-of-stars trajectory, 2) a brain dataset acquired with a ramp-sampled UTE radial trajectory, and 3) a cardiac dataset with a variable density spiral trajectory. We also applied on one 3D UTE dataset to illustrate the additional benefit of using preconditioners on 3D datasets. These datasets are described in more detail in Section \ref{sec:data}. \revised{\rnum{R1.6}The regularization parameters were manually selected to provide reconstructions with good subjective image quality.}

\revised{We consider three different k-space preconditioners to use with PDHG for comparisons: 1) \rnum{R2.2, R3.1}the Pipe-Menon density compensation factor~\cite{pipe1999} which was originally designed for gridding reconstruction, 2) single-channel diagonal k-space preconditioner, Eq.~\eqref{eq:sc_precond}, which will be abbreivated as SC k-space preconditioner, and 3) multi-channel diagonal k-space preconditioner Eq.~\eqref{eq:mc_precond}, which will be abbreivated as MC k-space preconditioner.}

For $\ell 2$-regularized reconstruction, conjugate gradient (CG) with and without circulant preconditioning, and PDHG with various k-space preconditionings were applied and compared with $\lambda = 0.01$. For $\ell 1$-wavelet regularized reconstruction, FISTA, \revised{\rnum{R3.3}ADMM with and without circulant preconditioning,} and PDHG with various k-space preconditionings were applied and compared with $\lambda=0.001$. The Daubechies-4 wavelet transform was used. For total variation regularized reconstruction, \revised{ADMM with and without circulant preconditioning, and PDHG with various k-space preconditionings} were applied and compared with $\lambda=0.001$. Anisotropic total variation along horizontal and vertical directions were used. \revisedthree{For ADMM, adaptive parameter selection as detailed in~\cite{Boyd:2011bw} was used for the convergence parameter $\rho$ with the initial value set to 1. CG with a tolerance of $1$ were used to solve the sub-problems in ADMM. The parameters for ADMM were selected by sweeping various values over the liver datasets without preconditioning as shown in Supplementary Figure~\ref{fig:l1_wavelet_admm_cg_tol} and~\ref{fig:l1_wavelet_admm_rho}. Similar behaviours were seen for ADMM with circulant preconditioning in Supplementary Figure~\ref{fig:l1_wavelet_admm_cp_cg_tol} and~\ref{fig:l1_wavelet_admm_cp_rho}.}

\revised{The circulant preconditioner used for CG and ADMM is designed with a least squares metric that takes into account of the sensitivity maps and the regularization function. For single-channel non-Cartesian imaging, the preconditioner coincides with Chan's preconditioner~\cite{chan1988} used in Weller et al.~\cite{weller2014}. And for multi-channel Cartesian imaging with total variation regularization, the preconditioner has \revisedtwo{similar structure as Koolstra et al's precondtioner}~\cite{koolstra2018}. Appendix~\ref{sec:appendix_circulant} contains the detailed derivation.}

All methods were implemented in Python using the software packages NumPy~\cite{walt2011} and CuPy~\cite{cupy2017} on a workstation with four Nvidia Titan Xp GPUs. All operations, except the wavelet transform, were run on a single GPU. \revised{\rnum{R1.9-R1.12}Normalized coil sensitivity maps were obtained using ESPIRiT~\cite{uecker2014} on the gridded low frequency k-space.} The NUFFT operations were implemented following Beatty et al.~\cite{beatty_rapid_2005} with an oversampling ratio of 1.25 and an interpolation kernel width of 4. \revised{\rnum{R3.2}The maximum eigenvalue $\lambda_{\text{max}}(\mathbf{P} \mathbf{A} \mathbf{A}^H)$ used for PDHG was estimated using the power method with 30 iterations.}

For the 2D datasets, all methods were run for 30 iterations (outer iterations for ADMM), and the objective values were computed for each iteration. For the 3D dataset, all methods were run for 1000 iterations. For $\ell 2$-regularized reconstructions, the computation time for constructing the circulant preconditioner and the proposed preconditioner was also recorded.

\revisedtwo{Finally, we note that algorithms compared in this work have subtle differences than those used in prior works. This is because each prior work considers a different reconstruction setting and it becomes impossible to evaluate each method in the same way. In particular, a monotonic version of FISTA was used in Ramani and Fessler~\cite{ramani2011}, which has higher computational complexity than the vanilla FISTA compared in this work. In addition, both Ramani and Fessler~\cite{ramani2011} and Weller et al.~\cite{weller2014} consider more complex regularization functions than the ones in this work. Koolstra et al. also considered Cartesian imaging with both $\ell 1$-wavelet and total-variation regularizations, which can exhibit very different convergence behaviors.} 

\subsection{Dataset Details}
\label{sec:data}

The liver dataset was acquired with a stack-of-stars trajectory using a 3D T1-FFE sequence (TR/TE 4.35 ms/1.20ms, resolution $1 \times 1 \times 1.5$ mm$^3$, field-of-view $40 \times 40 \times 12.5$ cm$^3$, \revised{\rnum{R1.9-R1.12}number of spokes $378$, reconstruction matrix size of $401 \times 401$ with an effective undersampling factor of about 1.66}). The sequence was implemented on a 3T MR system (Philips Healthcare) equipped with a 16-channel torso coil. The center slice was extracted after taking an inverse FFT along the slice direction for the experiments.

The cardiac dataset was acquired with a variable density spiral trajectory on a 1.5 T GE scanner (GE Healthcare, Waukesha, WI) with an 8-channel cardiac coil and the HeartVista RTHawk platform (HeartVista, Los Altos, CA). The trajectory consists of 3 interleaves with \revised{\rnum{R1.9-R1.12}3996 readout points and an effective undersampling factor of about 8.} It has an FOV of $32 \times 32$ cm$^2$, a reconstruction matrix size of $320 \times 320$ and TR of 25.8 ms.

The brain dataset was acquired with a centered-out radial trajectory on a 7.0 T GE clinical scanner (GE Healthcare, Waukesha, WI) with 8-channel head coil. The trajectory has 256 readout points and 754 half-spokes. The following prescribed parameters were used: flip angle of 5 degree, field-of-view $20 \times 20$ cm$^2$, in-plane resolution $1 \times 1$ mm$^2$, \revised{\rnum{R1.9-R1.12}reconstruction matrix size of $255 \times 200$}, and TE/TR = 3.4 ms/2 s.

The 3D UTE dataset was acquired on a \revised{\rnum{R1.9-R1.12}3 T GE clinical scanner (GE Healthcare, Waukesha, WI)} with 8-channel body coil with an optimized bit-reversed ordered radial trajectory~\cite{johnson2013optimized} using the sequence described in~\cite{jiang2018}. The following prescribed parameters were used: FOV of $32 \times 32 \times 32$ cm$^3$, \revised{\rnum{R1.9-R1.12}reconstruction matrix size of $356 \times 206 \times 211$ with an effective undersampling factor of about 3}, flip angle of 4 degrees, 1.25 mm isotropic resolution, sampling bandwidth of 62.5 kHz, and readout duration of 1 ms. 75,800 spokes were acquired.

\subsection{Results}
\label{sec:results}

Figure~\ref{fig:sense_brain} shows the iteration progression for the $\ell 2$-regularized reconstruction of the liver dataset. Both visually and quantitatively in terms of computation time, the proposed preconditioning methods (SC and MC) along with the vanilla CG converge faster than other methods. \revised{The per-iteration computation time for CG with circulant preconditioning is noticeably longer than for other methods, which results in slower convergence rate in terms of computation time. This is also supported by the Supplementary Figures~\ref{fig:sense_liver} and~\ref{fig:sense_cardiac}.}

\begin{figure}[!ht]
\begin{center}
  \includegraphics[width=\linewidth]{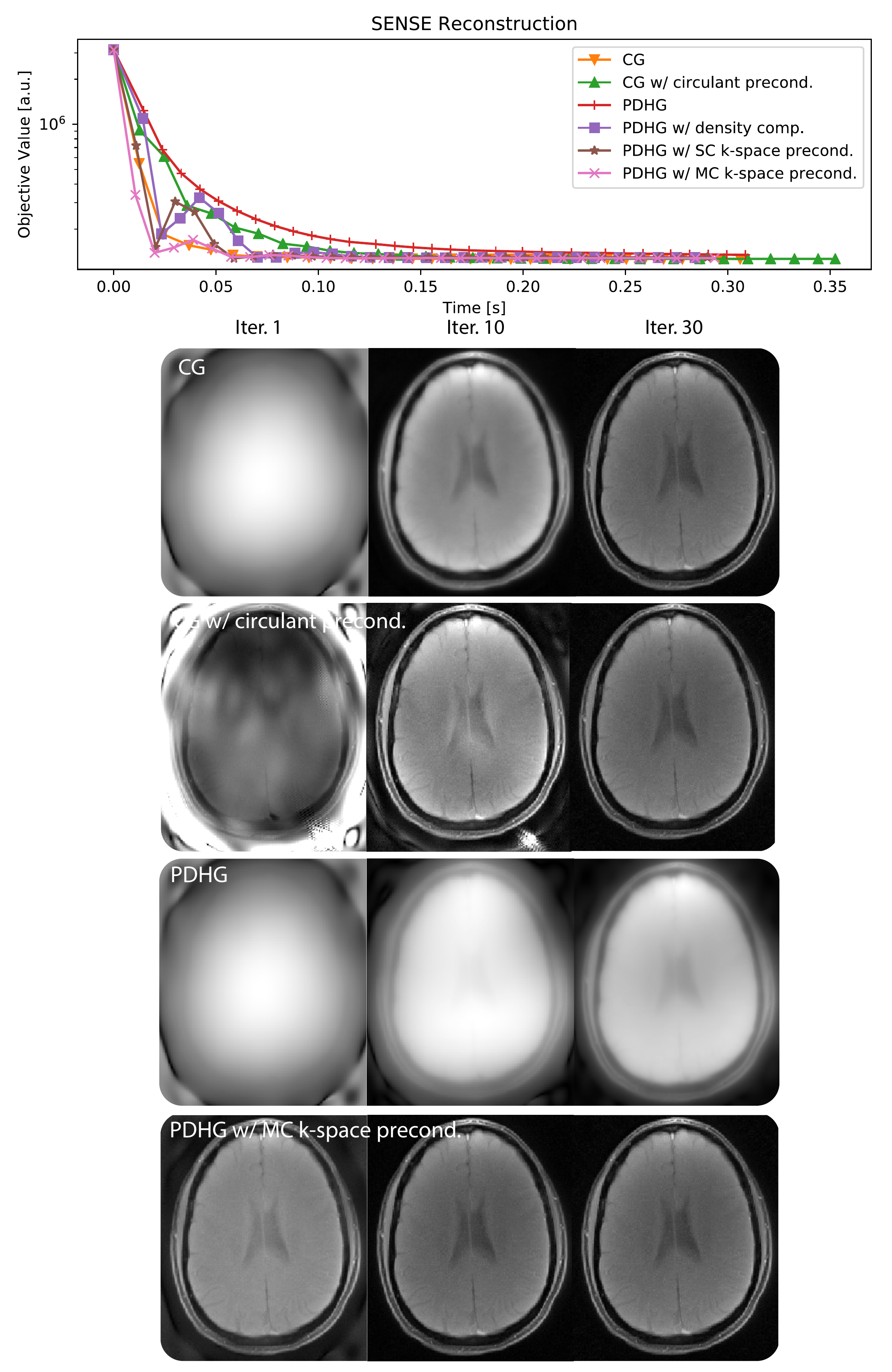}
\caption{Iteration progression for $\ell 2$ regularized reconstruction of the brain dataset. Each marker denotes the objective value calculated at each outer iteration. CG and the proposed preconditioning methods converge faster than other methods with respect to computation time.}
\label{fig:sense_brain}
\end{center}
\end{figure}

Table \ref{tab:preconditioner_construction} shows the computation time for constructing the preconditioners. The construction of the proposed preconditioner is about twice as slow as constructing the circulant preconditioner, but is still in a reasonable range. \revised{For applications in which the coil sensitivity maps are calculated from the pre-scan or estimated from the first scan and used multiple times for a sequence of scans, this overhead can be negligible. For applications in which the recalculation time matters, the SC kspace precondioner is preferred as it can be precomputed.}

\begin{table}[!ht]
  \centering
  \caption {Computation time for constructing preconditioners}
  \label{tab:preconditioner_construction} 
  \begin{tabular}{|l|l|l|l|}
    \hline
    & \textbf{Liver} & \textbf{Cardiac} & \textbf{Brain} \\ \hline
    \textbf{Circulant precond.} & 0.0974 s & 0.0502 s & 0.0147 s \\ \hline
    \textbf{MC k-space precond.}   & 0.231 s & 0.117 s  & 0.0334 s     \\ \hline
  \end{tabular}
\end{table}

Figure~\ref{fig:l1_wavelet_cardiac} shows the iteration progression for $\ell 1$-wavelet regularized reconstruction of the cardiac dataset. Both visually and quantitatively in terms of computation time, the proposed preconditioning methods (SC and MC) converge faster than other methods. \revised{\rnum{R1.1}The figure also shows that ADMM with circulant preconditioning is slower than ADMM without preconditioning, which can be due to the additional FFTs. While PDHG with density compensation accelerates convergence with respect to the vanilla PDHG, it also shows excessive oscillations.} Supplementary Figures~\ref{fig:l1_wavelet_brain} and~\ref{fig:l1_wavelet_liver} also support these observations.

\begin{figure}[!ht]
\begin{center}
  \includegraphics[width=\linewidth]{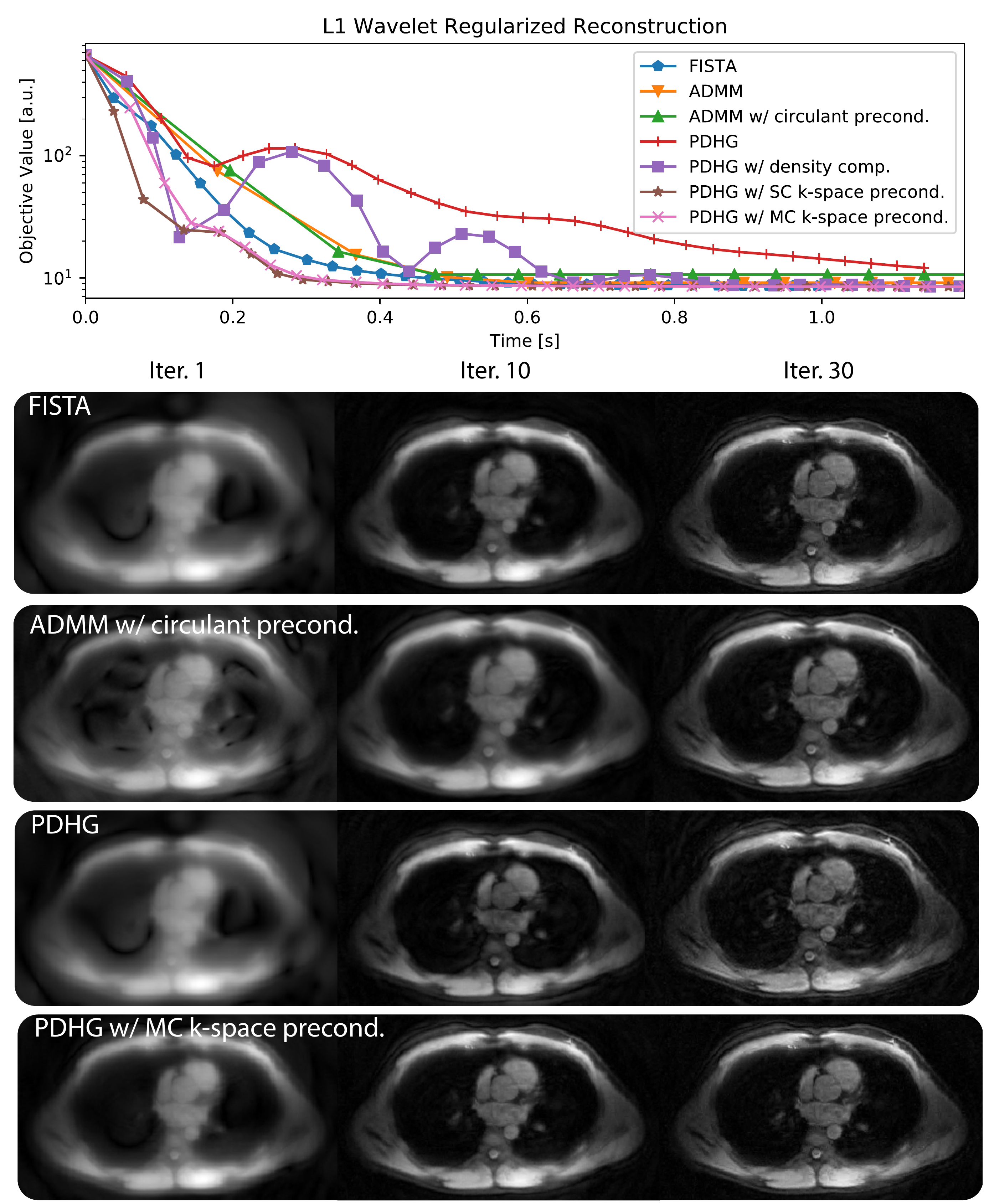}
\caption{Iteration progression for $\ell 1$ wavelet regularized reconstruction of the cardiac dataset. Each marker denotes the objective value calculated at each outer iteration. The proposed preconditioning methods converge faster than other methods with respect to computation time.}
\label{fig:l1_wavelet_cardiac}
\end{center}
\end{figure}

Figure~\ref{fig:total_variation_liver} shows the iteration progression for total variation regularized reconstruction of the liver dataset. Similar to the $\ell 1$-wavelet regularized reconstruction experiment, the proposed preconditioning methods converge faster than other methods. While ADMM with and without circulant preconditionings are competitive in terms of iteration number, their increased per-iteration time make them slightly slower in reaching an approximate optimal point. Other datasets shown in Supplementary Figures~\ref{fig:total_variation_brain} and~\ref{fig:total_variation_cardiac} support the above observations as well.

\begin{figure}[!ht]
\begin{center}
  \includegraphics[width=\linewidth]{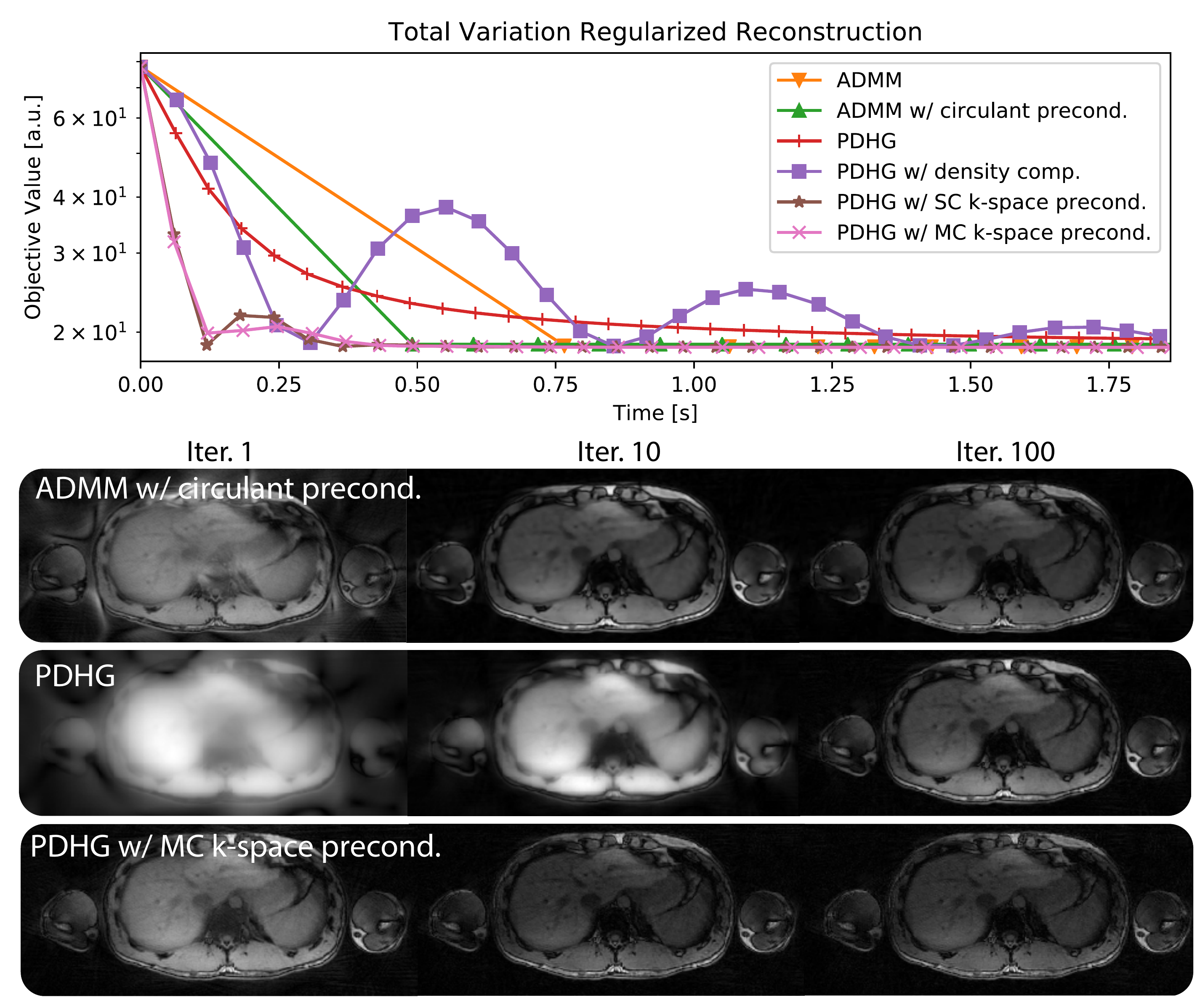}
\caption{Iteration progression for total variation regularized reconstruction of the liver dataset. Each marker denotes the objective value calculated at each outer iteration. The proposed preconditioning methods converge faster than other methods with respect to computation time.}
\label{fig:total_variation_liver}
\end{center}
\end{figure}

Finally, the iteration progression for the 3D UTE dataset was shown earlier in Figure~\ref{fig:l1_wavelet_uwute}. Both FISTA and PDHG exhibit extreme blurring even after 100 iterations. In contrast, PDHG with the proposed preconditioner converges in about ten iterations, both visually and quantitatively in terms of minimizing the objective value. This shows that the proposed method can offer an order magnitude speedup in 3D than in 2D.

\section{Discussion}
\label{sec:discussion}

In this article, we presented a preconditioning method using the convex dual formulation. This enables the use of efficient k-space operations as preconditioners and does not modify the objective function. Through experiments, we have demonstrated that the proposed technique indeed accelerates the convergence of non-Cartesian reconstructions.

In particular, we compared the performance of the proposed preconditioning to circulant preconditioning with CG for $\ell 2$-regularized reconstructions and with ADMM for $\ell 1$ wavelet and total variation regularized reconstructions. The main advantage of the proposed preconditioning lies in the per-iteration computation time. k-space preconditioning is faster than circulant preconditioning, and adds very little computational overhead. This is expected as the circulant preconditioning requires two additional FFTs per iteration, whereas the proposed k-space diagonal preconditioning requires only element-wise multiplications. \revised{Moreover, in our experiments, ADMM always incurs additional computational overhead compared to other methods without inner loops. We conjecture that this may be due to additional memory copying and variable initialization in the inner loops. For example, residual terms in CG need to be initialized before each inner loop.} \revisedtwo{On the other hand, we note that the computational disadvantage of inner loops can be reduced by pre-calculating the residual outside CG based on the solution found in the previous outer iteration.}

\revised{Although the MC k-space preconditioner achieves the fastest convergence most of the time, the SC k-space preconditioner provides similar or sometimes even faster convergence acceleration. Since the MC k-space preconditioner cannot be precomputed, we recommend using the SC k-space preconditioner for general settings. In applications in which the same sensitivity maps and trajectory are used repeatedly, then the MC k-space preconditioner can provide a slightly faster convergence behavior with less oscillation. We have also investigated using the Pipe-Menon density compensation factor as a k-space preconditioner. While in some cases it accelerates convergence, Figures~\ref{fig:l1_wavelet_cardiac} and~\ref{fig:total_variation_liver}, and Supplementary Figures~\ref{fig:sense_cardiac},~\ref{fig:total_variation_cardiac}, and~\ref{fig:total_variation_brain} show that it can introduce excessive oscillations, which slow down the overall convergence. Based on this, we recommend the SC k-space preconditioner over density compensation factors as k-space preconditioners.}

Finally, the experiment with the 3D UTE dataset in Figure~\ref{fig:l1_wavelet_uwute} shows that the method offers orders of magnitude speedup for 3D datasets. This is expected because 3D trajectories have a higher variation in k-space density than 2D trajectories. In particular, the proposed method converged in about ten iterations, whereas other methods did not, even after hundreds of iterations. We note that most experiments in this article are done on 2D datasets because of their fast evaluation time. In terms of practical application, we expect the proposed preconditioning methods to be vastly more effective and useful for 3D non-Cartesian imaging.

\section{Conclusion}

We have shown a method to speed up non-Cartesian iterative reconstruction that retains the per-iteration computational efficiency of density compensation and reconstruction accuracy of preconditioning methods. In contrast to most existing preconditioning methods, the proposed technique does not increase the per-iteration computation time compared to vanilla iterative methods, such as the conjugate gradient method. With the proposed preconditioning, iterative reconstruction for non-Cartesian imaging can be accelerated without compromises.
\begin{appendices}

\section{Derivation for the $\ell 2$ optimized diagonal preconditioner}
\label{sec:appendix_precond}

We are interested in solving the following minimization problem:
\begin{align*}
  \min_{\mathbf{p}} \frac{1}{2} \left\|  \text{diag}(\mathbf{p}) \mathbf{A} \mathbf{A}^H - \mathbf{I} \right\|_F^2
\end{align*}

Expanding the objective function element-by-element, we obtain
\begin{align*}
  \min_{\vp} \frac{1}{2} \sum_{i = 1}^M \sum_{j = 1}^M \left| \mathbf{p}[i] \mathbf{a}_i^H \mathbf{a}_j - \bm{\delta}_{ij}\right|^2,
\end{align*}
where $\bm{\delta}$ is the Dirac delta function.

Taking the gradient with respect to $\vp_i$, setting it to zero and re-arranging, we have,
\begin{align*}
  {\mathbf{p}[i]}^{-1} = \frac{ \sum_{j=1}^M |\mathbf{a}_i^H \mathbf{a}_j|^2 } { \|\mathbf{a}_i\|_2^2 }.
\end{align*}

\section{\revised{\rnum{R1.4, R1.7}Derivation for the $\ell 2$ optimized Circulant Preconditioner}}
\label{sec:appendix_circulant}

The circulant preconditioner we consider in this article minimizes the following problem:
\begin{align*}
  \mathbf{P} = \underset{\mathbf{P}~\text{circulant}}{\text{argmin}} \left\|  \mathbf{A}^H \mathbf{A} - \mathbf{P} \right\|_F^2.
\end{align*}

Each element of $\mathbf{A}^H \mathbf{A}$ is given by,
\begin{align*}
    (\mathbf{A}^H \mathbf{A})[m, n] = \frac{1}{N} \sum_{i = 1}^M \sum_{c = 1}^C \mathbf{s}^*_{c}[m] \mathbf{s}_{c}[n] e^{\imath 2 \pi f_i (m - n)/N}
\end{align*}
and the circulant matrix $\mathbf{P}$ has the form,
\begin{align*}
    \mathbf{P}[m, n] = \mathbf{p}[{((m - n))_N}],
\end{align*}
where $\mathbf{p} \in \mathbb{C}^N$ is the underlying convolution kernel, and $((~))_N$ denotes the modulo operation by $N$.

Hence, minimizing the least squares criterion results in,
\begin{align*}
    \mathbf{p}[k] &= \frac{1}{N} \sum_{((m - n))_N = k}  (\mathbf{A}^H \mathbf{A})[m, n].
\end{align*}

Let us define the autocorrelation function $\mathbf{r} \in \mathbb{C}^{2N - 1}$ and the point-spread function $\mathbf{h} \in \mathbb{C}^{2N - 1}$ as follows,
\begin{align*}
 \mathbf{r}[k] &= \frac{1}{\sqrt{N}} \sum_{c = 1}^C \sum_{m - n = k} \mathbf{s}^*_{c}[m] \mathbf{s}_{c}[n] \\
 \mathbf{h}[k] &= \frac{1}{\sqrt{N}}\sum_{i = 1}^M e^{\imath 2 \pi f_i k/N},
\end{align*}
then after some algebra, the convolution kernel can be expressed as,
\begin{align*}
    \mathbf{p}[k] = \mathbf{r}[k] \mathbf{h}[k] + \mathbf{r}[k - N] \mathbf{h}[k - N].
\end{align*}

\end{appendices}

\printbibliography
\endgroup

\clearpage
\begin{center}
\Large\textbf{Supporting Materials}
\end{center}
   
\renewcommand\thesuppfigure{S\arabic{suppfigure}}  

\section*{Derivation for the Convex Dual Problem}
\label{sec:appendix_dual}

Here we will derive the dual problem through the Lagrangian function. Let us first introduce a variable $\mathbf{z}$ to make the objective function~\eqref{eq:obj} a constrained optimization problem:
\begin{align*}
  &\min_{\vx, \vz} \frac{1}{2}\| \vz - \vy \|_2^2 + g(\vx) \\
  &\text{subject to:}\quad \vz = \mA\vx
\end{align*}

Introducing a Lagrangian variable $\mathbf{u}$ gives us,
\begin{align*}
  \min_{\vx, \vz} \max_\mathbf{u} \frac{1}{2}\| \vz - \vy \|_2^2 + g(\vx) + \Re \langle \mathbf{u}, \vz - \mA\vx \rangle
\end{align*}

Switching the min and the max, and minimizing over $\mathbf{z}$ gives us $\vz = \vy - \mathbf{u}$. Substituting and re-arranging, we obtain,
\begin{align*}
 \max_\mathbf{u} \min_{\vx} - \frac{1}{2}\| \mathbf{u}\|_2^2 + \Re \langle \mathbf{u}, \vy \rangle  + g(\vx) -  \Re \langle \mA^H \mathbf{u}, \vx \rangle.
\end{align*}

Using the definition of the conjugate function $g^*(\vx^*) = \max_{\vx^*} \langle \vx^*, \vx\rangle - g(\vx)$, we have,
\begin{align*}
 \max_\mathbf{u}  - \frac{1}{2}\| \mathbf{u}\|_2^2 + \Re \langle \mathbf{u}, \vy \rangle  - g^*(-\mA^H \mathbf{u}).
\end{align*}

\section*{\revised{\rnum{R1.2}Connection to Trzasko et al.~\cite{trzasko2014}}}
\label{sec:appendix_trzakso}

Here we re-derive the result in Trzasko et al. through convex duality. Let us consider $g(\mathbf{x}) = \frac{\lambda}{2} \| \mathbf{x} \|_2^2$, then the dual problem is given by,
\begin{align*}
  \max_\mathbf{u} -\left( \frac{1}{2}\|\mathbf{u}\|_2^2 - \Re \langle \mathbf{u}, \mathbf{y} \rangle + \frac{1}{2 \lambda}\|\mathbf{A}^H \mathbf{u}\|_2^2 \right),
\end{align*}
which has the optimality condition
\begin{align*}
  (\mathbf{A} \mathbf{A}^H + \lambda \mathbf{I}) \mathbf{u} = \lambda \mathbf{y}.
\end{align*}
Hence, we can precondition in k-space by preconditioning the dual variable by solving:
\begin{align*}
  \mathbf{P} (\mathbf{A} \mathbf{A}^H + \lambda \mathbf{I}) \mathbf{u} = \lambda \mathbf{P} \mathbf{y}
\end{align*}

In general, the primal and dual variables $\mathbf{x}$ and $\mathbf{u}$ are connected with the following relationship:
\begin{equation}
  \begin{aligned}
  - \mathbf{A}^H \mathbf{u} &\in \partial g (\mathbf{x}) \\
  \mathbf{A} \mathbf{x} &= \mathbf{u} + \mathbf{y},
  \end{aligned}
  \label{eq:connection}
\end{equation}
where $\partial g(\mathbf{x})$ denotes the sub-differential of $g$ at $\mathbf{x}$.

Since $\partial g(\mathbf{x}) = \{\lambda \mathbf{x}\}$, from the primal dual relationship~\eqref{eq:connection} we can recover the primal variable by performing,
\begin{align*}
  \mathbf{x} = \frac{1}{\lambda} \mathbf{A}^H \mathbf{u}.
\end{align*}

The above method is precisely what Trzasko et al.~\cite{trzasko2014} proposed for the $\ell 2$-regularized sub-problem within ADMM.

\section*{Supporting Figures}
\begin{suppfigure}[!ht]
\begin{center}
  \includegraphics[width=\linewidth]{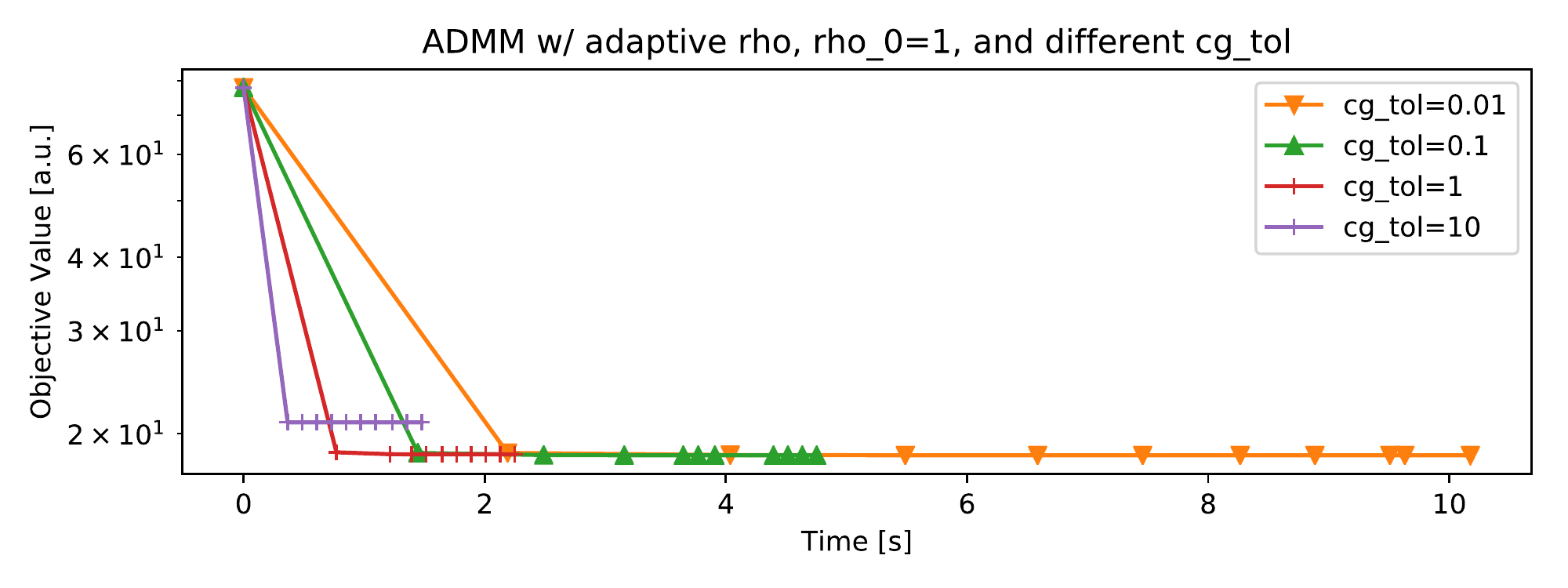}
\caption{\revisedthree{Iteration progression over different ADMM CG tolerance for $\ell 1$-wavelet regularized reconstruction of the liver dataset.}}
\label{fig:l1_wavelet_admm_cg_tol}
\end{center}
\end{suppfigure}

\begin{suppfigure}[!ht]
\begin{center}
  \includegraphics[width=\linewidth]{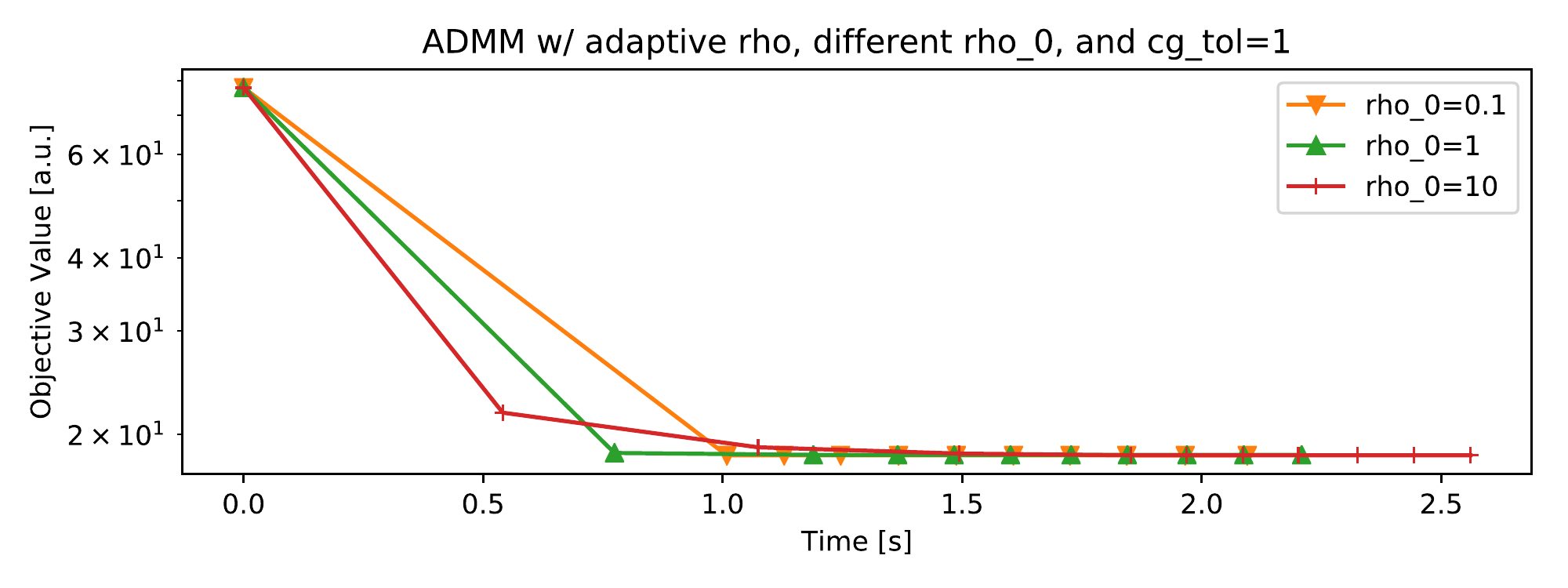}
\caption{\revisedthree{Iteration progression over different ADMM convergence parameter $\rho$ for $\ell 1$-wavelet regularized reconstruction of the liver dataset.}}
\label{fig:l1_wavelet_admm_rho}
\end{center}
\end{suppfigure}

\begin{suppfigure}[!ht]
\begin{center}
  \includegraphics[width=\linewidth]{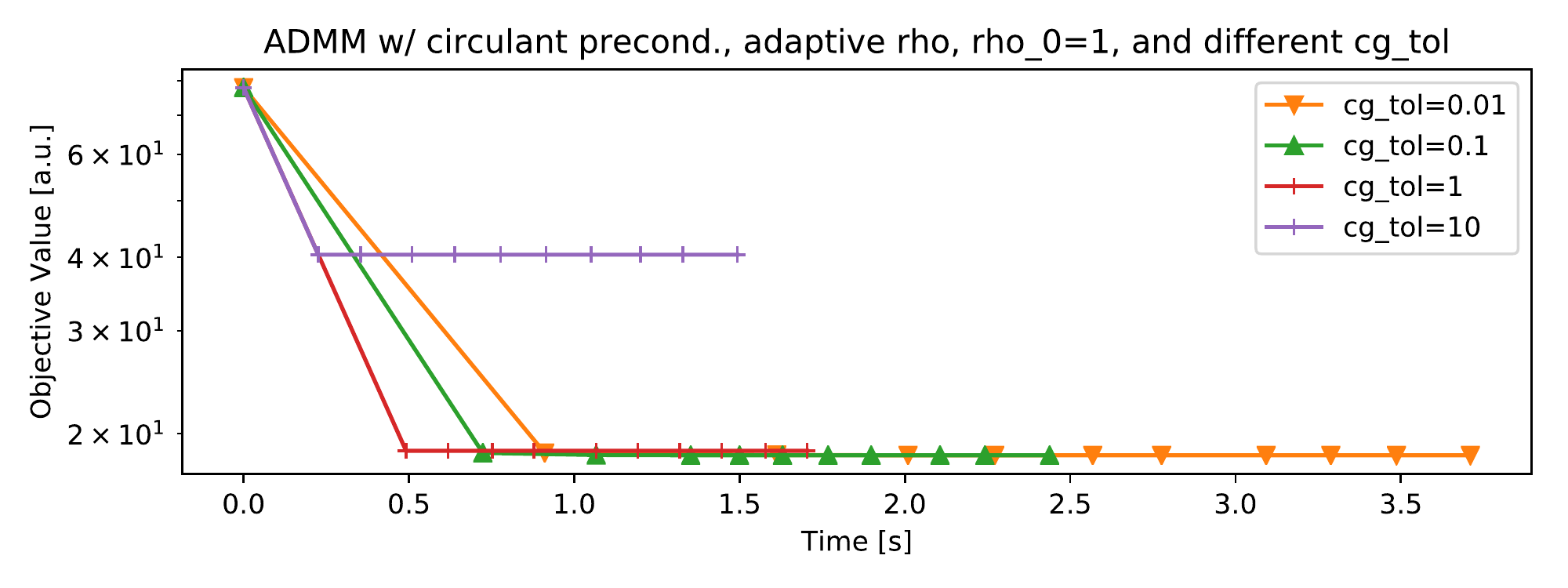}
\caption{\revisedthree{Iteration progression over different ADMM CG tolerance with preconditioning for $\ell 1$-wavelet regularized reconstruction of the liver dataset.}}
\label{fig:l1_wavelet_admm_cp_cg_tol}
\end{center}
\end{suppfigure}

\begin{suppfigure}[!ht]
\begin{center}
  \includegraphics[width=\linewidth]{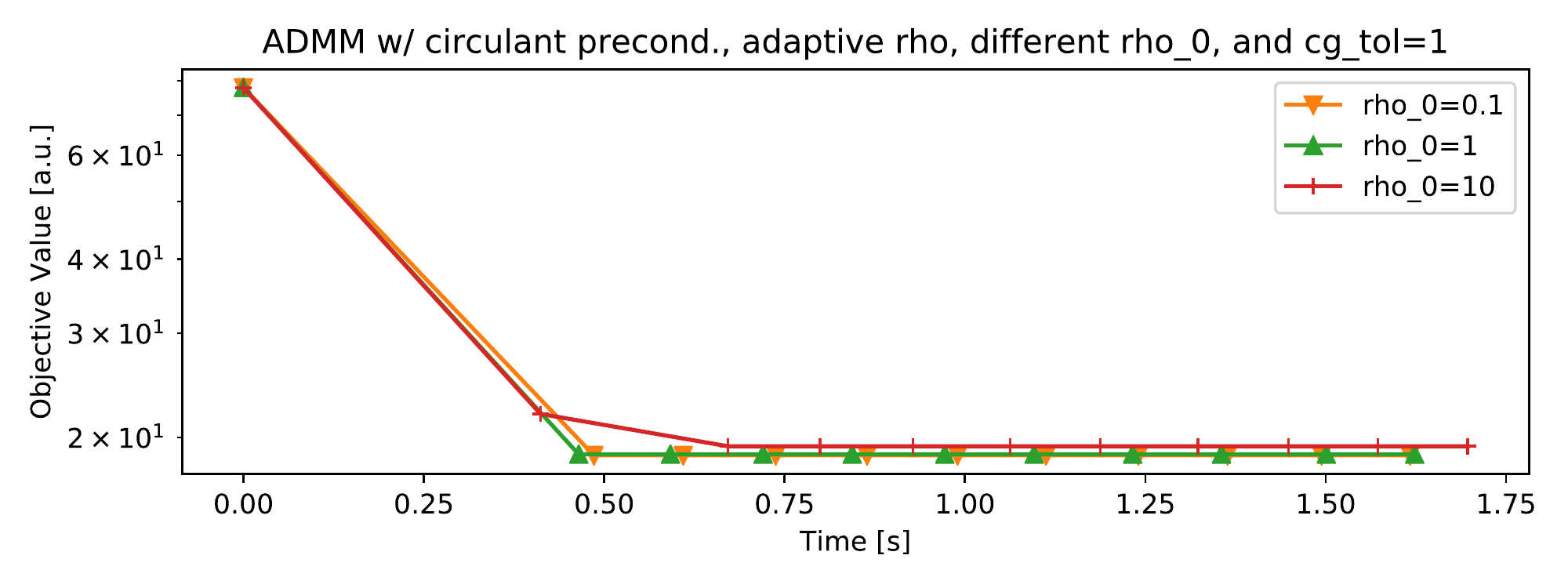}
\caption{\revisedthree{Iteration progression over different ADMM convergence parameter $\rho$ with preconditioning for $\ell 1$-wavelet regularized reconstruction of the liver dataset.}}
\label{fig:l1_wavelet_admm_cp_rho}
\end{center}
\end{suppfigure}

\begin{suppfigure}[!ht]
\begin{center}
  \includegraphics[width=\linewidth]{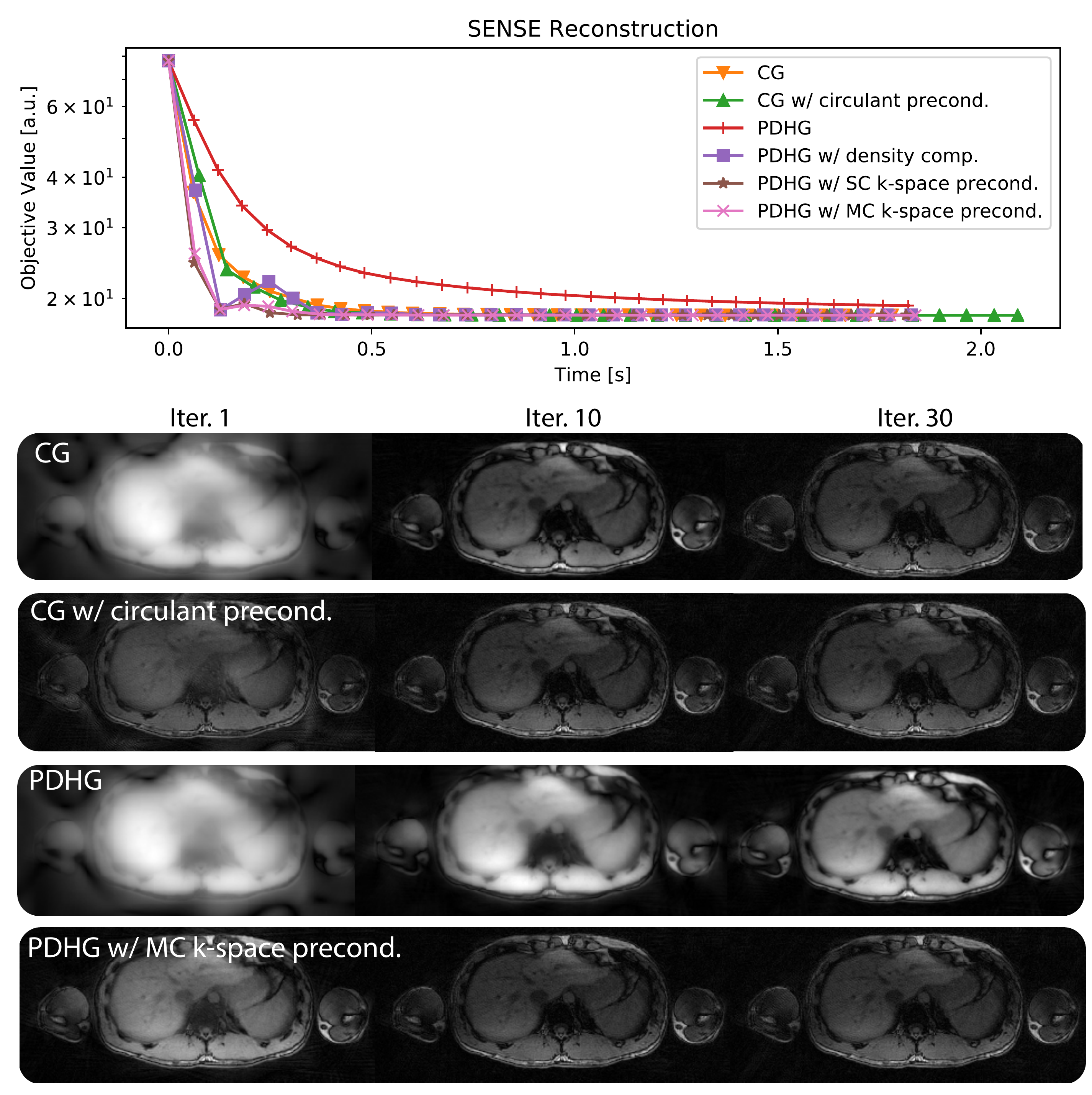}
\caption{Iteration progression for $\ell 2$-regularized reconstruction of the liver dataset.}
\label{fig:sense_liver}
\end{center}
\end{suppfigure}

\begin{suppfigure}[!ht]
\begin{center}
  \includegraphics[width=\linewidth]{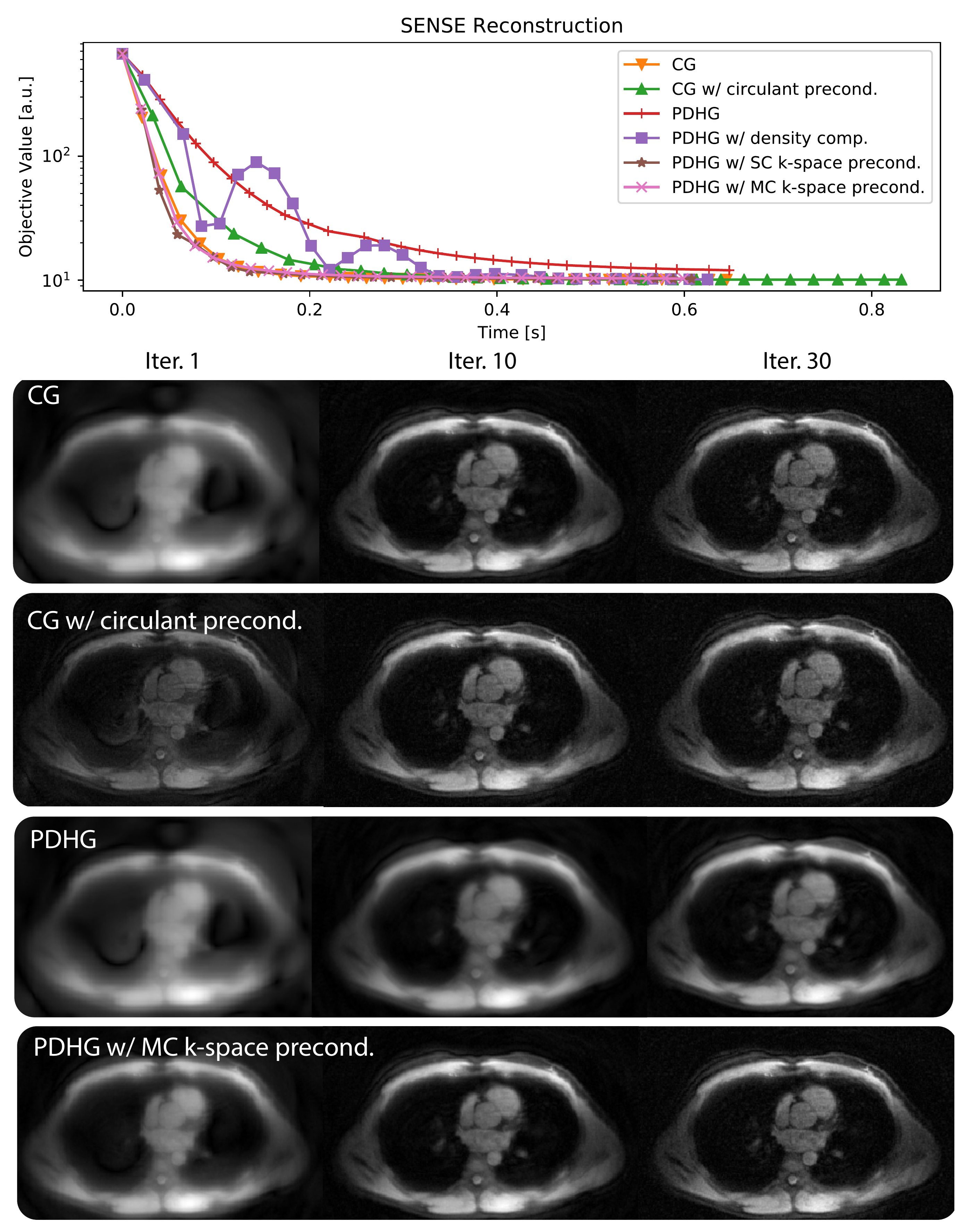}
\caption{Iteration progression for $\ell 2$-regularized reconstruction of the cardiac dataset.}
\label{fig:sense_cardiac}
\end{center}
\end{suppfigure}

\begin{suppfigure}[!ht]
\begin{center}
  \includegraphics[width=\linewidth]{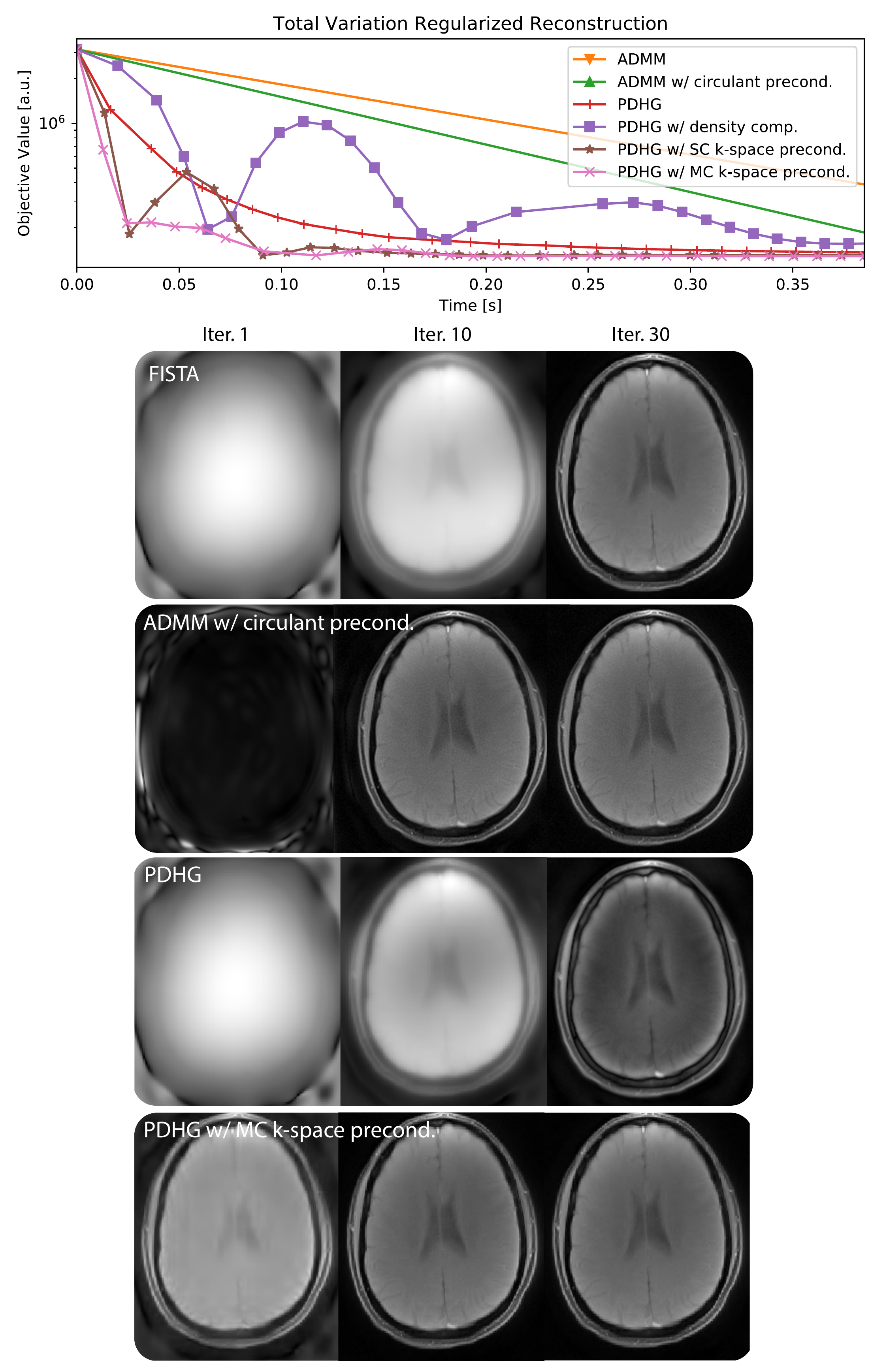}
\caption{Iteration progression for $\ell 1$ wavelet regularized reconstruction of the brain dataset.}
\label{fig:l1_wavelet_brain}
\end{center}
\end{suppfigure}

\begin{suppfigure}[!ht]
\begin{center}
  \includegraphics[width=\linewidth]{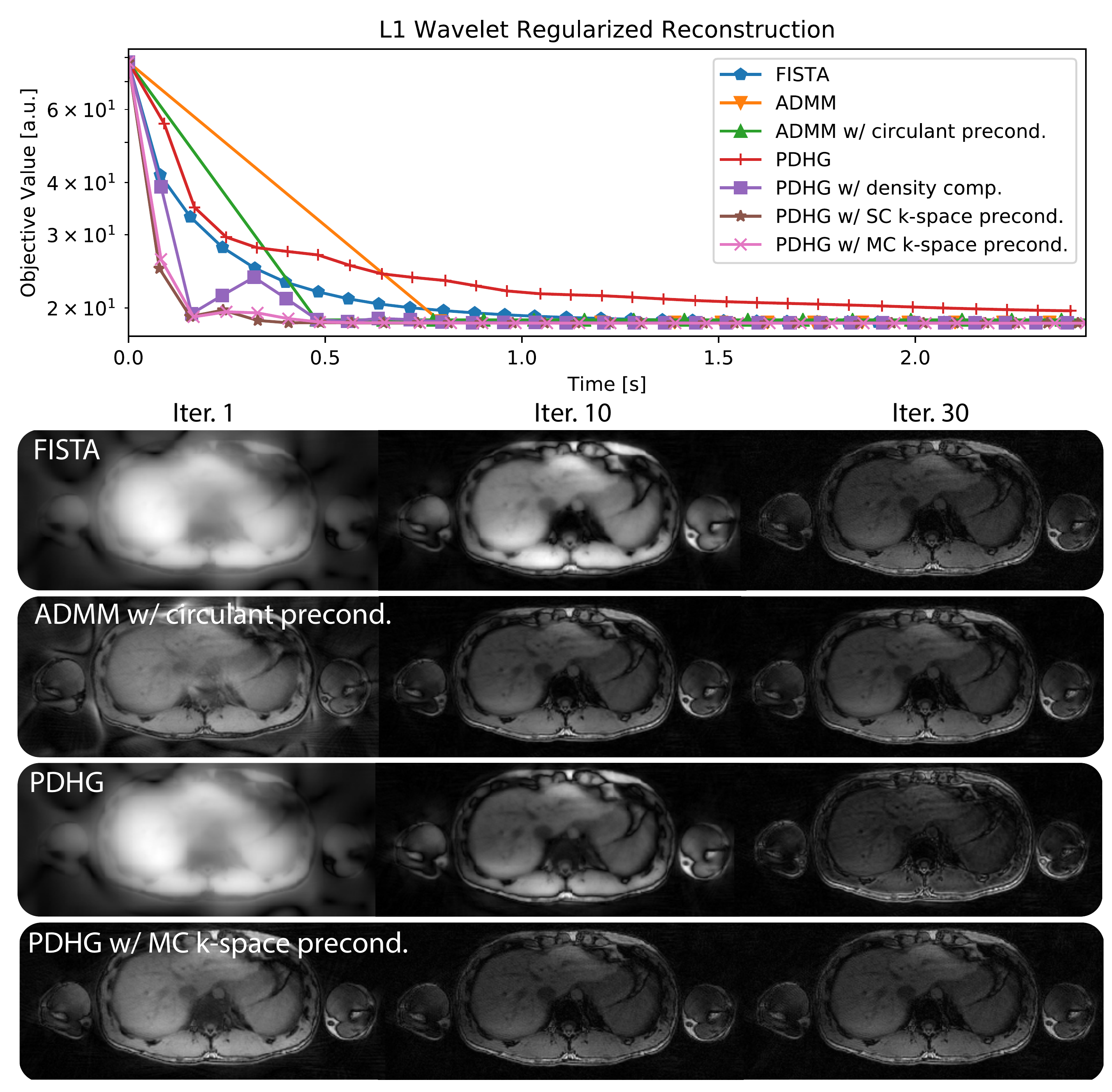}
\caption{Iteration progression for $\ell 1$ wavelet regularized reconstruction of the liver dataset.}
\label{fig:l1_wavelet_liver}
\end{center}
\end{suppfigure}

\begin{suppfigure}[!ht]
\begin{center}
  \includegraphics[width=\linewidth]{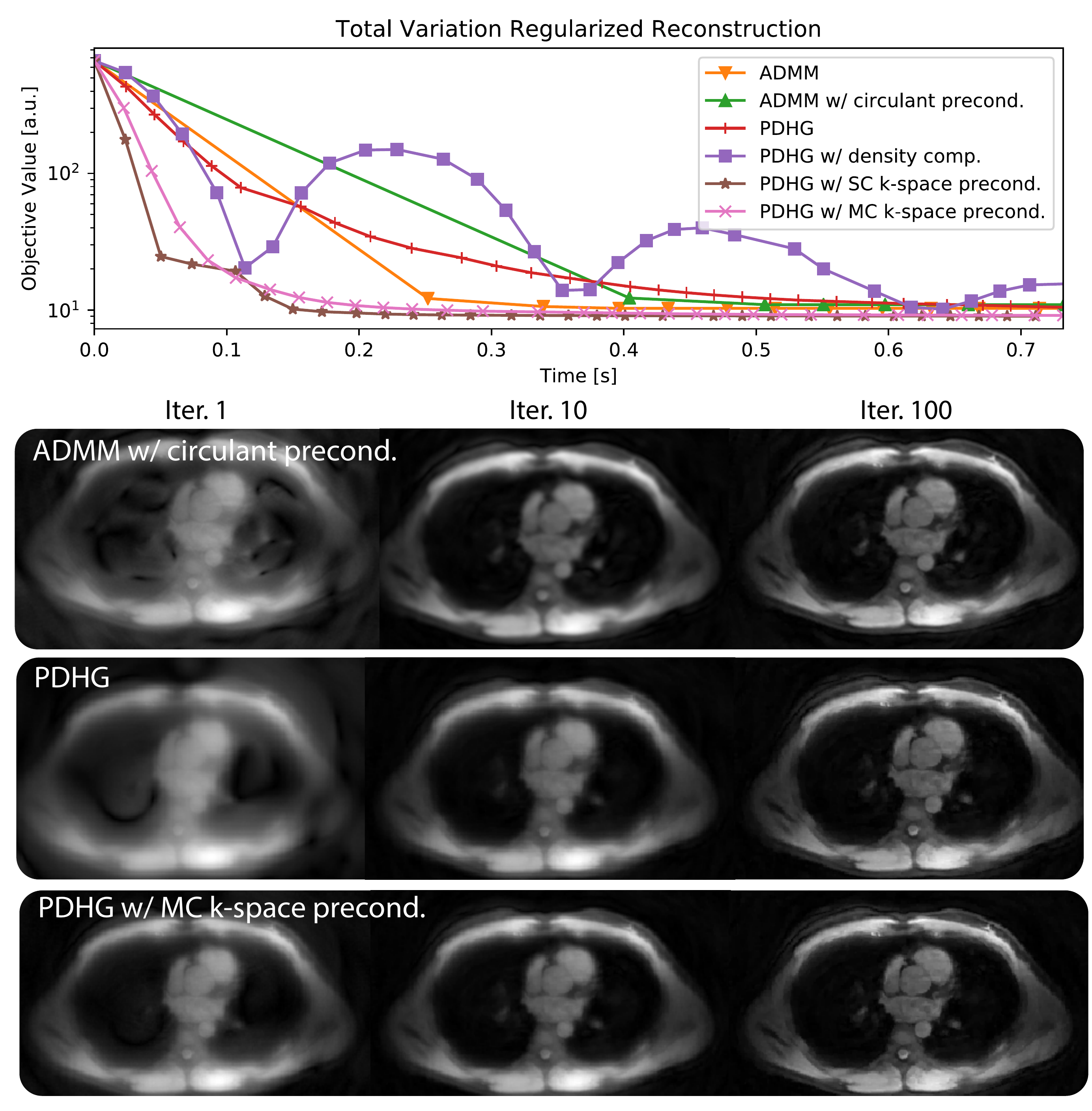}
\caption{Iteration progression for total variation regularized reconstruction of the cardiac dataset.}
\label{fig:total_variation_cardiac}
\end{center}
\end{suppfigure}

\begin{suppfigure}[!ht]
\begin{center}
  \includegraphics[width=\linewidth]{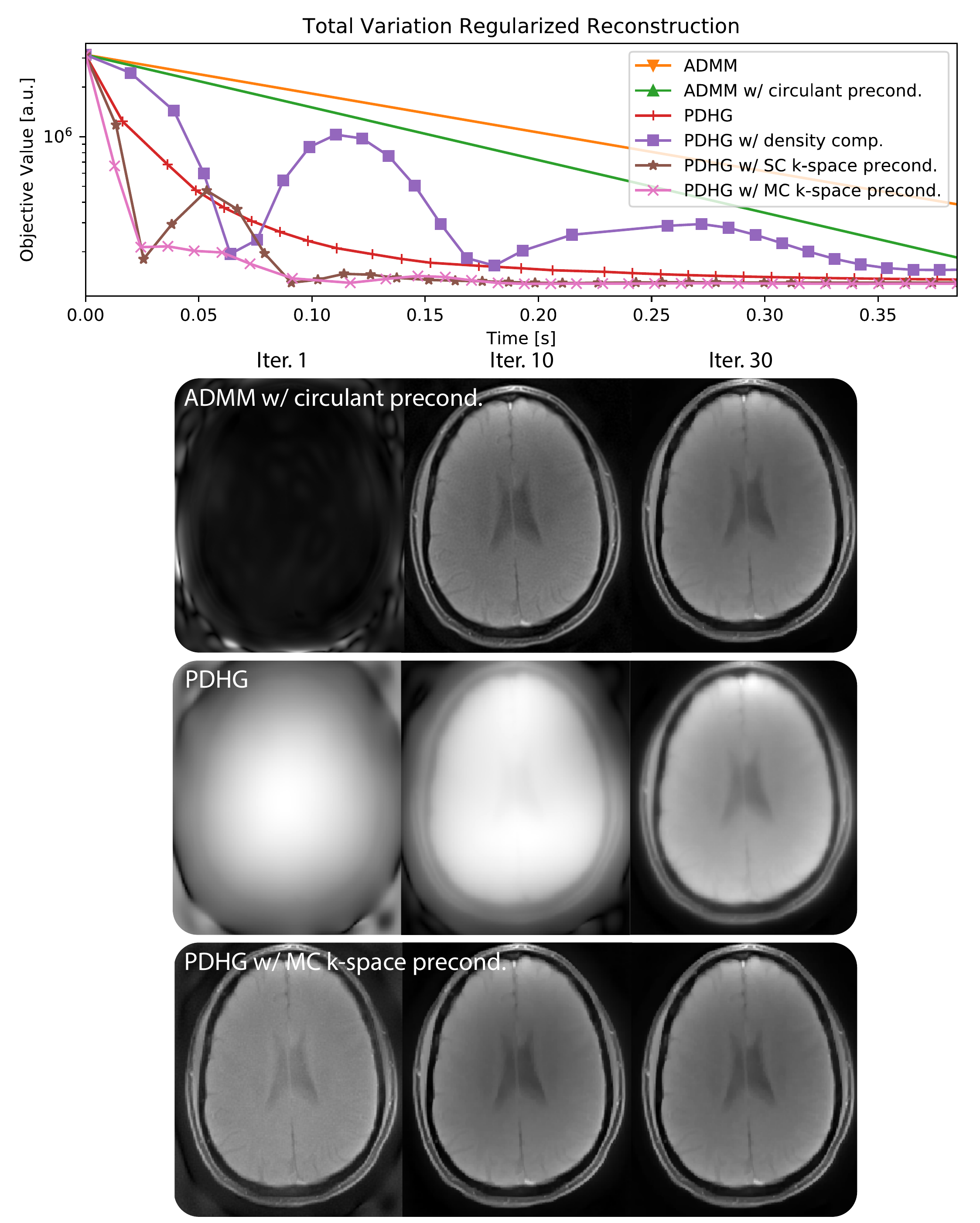}
\caption{Iteration progression for total variation regularized reconstruction of the brain dataset.}
\label{fig:total_variation_brain}
\end{center}
\end{suppfigure}

\end{document}